\documentstyle[12pt,epsf]{article}
\renewcommand{\theequation}{\thesection.\arabic{equation}}
\newlength{\extraspace}
\newlength{\extraspaces}
\newcommand{\be}{\begin{equation}
\addtolength{\abovedisplayskip}{\extraspaces}
\addtolength{\belowdisplayskip}{\extraspaces}
\addtolength{\abovedisplayshortskip}{\extraspace}
\addtolength{\belowdisplayshortskip}{\extraspace}}
\newcommand{\ee}{\end{equation}}
 
\newcommand{\ba}{\begin{eqnarray}
\addtolength{\abovedisplayskip}{\extraspaces}
\addtolength{\belowdisplayskip}{\extraspaces}
\addtolength{\abovedisplayshortskip}{\extraspace}
\addtolength{\belowdisplayshortskip}{\extraspace}}
\newcommand{\ea}{\end{eqnarray}}

\newcommand{\bas}{\begin{eqnarray*}
\addtolength{\abovedisplayskip}{\extraspaces}
\addtolength{\belowdisplayskip}{\extraspaces}
\addtolength{\abovedisplayshortskip}{\extraspace}
\addtolength{\belowdisplayshortskip}{\extraspace}}
\newcommand{\eas}{\end{eqnarray*}}
 
\newcounter{subequation}[equation]
\makeatletter

\expandafter\let\expandafter
\reset@font\csname reset@font\endcsname

\def\subeqnarray{\arraycolsep1pt
    \def\@eqnnum\stepcounter##1{\stepcounter{subequation}%
        {\reset@font\rm(\theequation\alph{subequation})}}
\jot5mm     \eqnarray}

\makeatother

\newcommand{\newappendix}[1]{
\vspace{15mm}
\pagebreak[3]
\addtocounter{section}{1}
\setcounter{equation}{0}
\setcounter{subsection}{0}
\setcounter{footnote}{0}
\renewcommand{\theequation}{\Alph{section}.\arabic{equation}}
\begin{flushleft}
{\large\bf Appendix: #1}
\end{flushleft}
\nopagebreak
\medskip
\nopagebreak}

\newcommand{\newsection}[1]{
\vspace{15mm}
\pagebreak[3]
\addtocounter{section}{1}
\setcounter{equation}{0}
\setcounter{subsection}{0}
\setcounter{footnote}{0}
 
\begin{flushleft}
{\large\bf \thesection. #1}
\end{flushleft}
\nopagebreak
\medskip
\nopagebreak}
 
\newcommand{\newsubsection}[1]{
\vspace{1cm}
\pagebreak[3]
 
\addtocounter{subsection}{1}
\noindent{ \bf \thesection.\arabic{subsection} #1}
\nopagebreak
\vspace{2mm}
\nopagebreak}

\newcommand{\NP}[1]{Nucl.\ Phys.\ {\bf #1}}
\newcommand{\PL}[1]{Phys.\ Lett.\ {\bf #1}}
\newcommand{\CMP}[1]{Comm.\ Math.\ Phys.\ {\bf #1}}
\newcommand{\PR}[1]{Phys.\ Rev.\ {\bf #1}}
\newcommand{\PRL}[1]{Phys.\ Rev.\ Lett.\ {\bf #1}}

\newcommand{\C}{\mbox{$\,${\sf I}\hspace{-1.2ex}{\bf C}}}

\newcommand{\R}{\mbox{\rm I\hspace{-.4ex}R}}
\newcommand{\1}{\mbox{1\hspace{-.6ex}1}}
\newcommand{\bra}{\langle}
\newcommand{\ket}{\rangle}
\newcommand{\ra}{\rightarrow}

\newcommand{\rra}{\ \longrightarrow \ }

\newcommand{\is}{ &\!=\!& }
\newcommand{\nonum}{\nonumber \\[1.5mm]}
\newcommand{\sspace}{\makebox[1cm]{ }}
\newcommand{\bspace}{\makebox[2cm]{ }}
\newcommand{\nspace}{\!\!\!\!\!\!\!\!\!\!}

\newcommand{\Tr}{{\rm Tr}}

\newcommand{\inv}{^{-1}}

\newcommand{\th}{{\theta}}
\newcommand{\lb}{\lambda}
\newcommand{\sh}{{\rm sh}}
\newcommand{\ch}{{\rm ch}}
\renewcommand{\d}{{\delta}}
\newcommand{\dd}[1]{ \frac{\partial}{\partial{#1}} }

\newcommand{\Fbar}{{\overline{F}}}

\newcommand{\cK}{{\cal K}}

\newcommand{\cN}{{\cal N}}
\newcommand{\cO}{{\cal O}}

\newcommand{\s}{\sigma}

\newcommand{\g}{\gamma}

\newcommand{\num}{{\rm num}}
\newcommand{\den}{{\rm den}}

\begin{document}
%
\begin{titlepage}
%
%
\begin{flushright}
AEI-068
\end{flushright}
\vspace{15mm}
 
\begin{center}
{\LARGE Varying the Unruh Temperature in}\\[3mm] 
{\LARGE Integrable Quantum Field Theories}\\[3cm]
{\large Max Niedermaier}\\ [3mm]
{\small\sl Max-Planck-Institut f\"{u}r Gravitationsphysik} \\
{\small\sl (Albert-Einstein-Institut)} \\
{\small\sl Schlaatzweg 1, D-14473 Potsdam, Germany}
\vspace{3.5cm}
 
{\bf Abstract}
\end{center}

\begin{quote}
A computational scheme is developed to determine the response
of a quantum field theory (QFT) with a factorized scattering 
operator under a variation of the Unruh temperature. 
To this end a new family of integrable systems is introduced, 
obtained by deforming such QFTs in a way that preserves the bootstrap
S-matrix. The deformation parameter $\beta$ plays the role of an 
inverse temperature for the thermal equilibrium states associated
with the Rindler wedge, $\beta = 2\pi$ being the QFT value. 
The form factor approach provides an explicit computational 
scheme for the $\beta \neq 2\pi$ systems, enforcing in particular a 
modification of the underlying kinematical arena. As examples 
deformed counterparts of the Ising model and the Sinh-Gordon model are 
considered.   
\end{quote}
\vfill

\renewcommand{\thefootnote}{\arabic{footnote}}
\setcounter{footnote}{0}
\end{titlepage}
\newsection{Introduction}

Sometimes it is advantageous to step outside ``flat land'' 
Minkowski space quantum field theory (QFT) even if the quantity 
one is interested in concerns the latter. A good example is the 
conformal anomaly. In a $2n$-dimensional (flat space) QFT it can be 
defined through a coefficient of an $n+1$-point function of the 
energy momentum tensor. Technically however it is useful to
first couple the system to some curved background and then 
compute the conformal anomaly as a suitable response 
with respect to a variation of the background metric (thereby 
producing the vacuum expectation value of the trace of the energy 
momentum tensor). Apart from the technical advantage that one only 
has to compute a 1-point function (in curved background) the result  
provides a linkage between flat space and curved space QFT.

\newsubsection{Thermalization and Replica}

Here we shall address the problem how to compute a similar 
response under a variation of the Unruh temperature. 
The latter refers to the well-known thermalization phenomenon
that the vacuum of a Minkowski space QFT `looks like' a
thermal state of inverse temperature $\beta = 2\pi$ (in natural
units) with respect to the Killing time of the Rindler wedge
\cite{Unruh,BW}. Heuristically one can summarize the result in
the symbolic identity
\be
\bra 0| \cO_1(x_1) \ldots  \cO_n(x_n) |0\ket =
\Tr[e^{2\pi K}  \cO_1(x_1) \ldots  \cO_n(x_n)]\;,
\;\;\;\mbox{if}\;x_1,\ldots, x_n \in W ,
\label{I1}
\ee
where $W$ is the right Rindler wedge, $K$ is the generator of 
Lorentz boosts in $W$ and $\cO_j(x)$ are some local quantum fields. 
If we momentarily ignore the fact that the trace can never exist%
\footnote{The spectrum of $K$ consists of the entire real line
so that $e^{2\pi K}$ is an unbounded operator. Conversely if 
$e^{2\pi K}$ were a positive trace-class operator (i.e. a 
density matrix) the spectrum of $-K$ necessarily would 
have to be discrete and bounded from below. Likewise a decomposition
of $K$ into a difference of left and right Rindler Hamiltonians does 
not really exist.} 
it is also clear what kind of operation one would like to perform, 
namely
\be
\beta \dd{\beta}\,\Tr[e^{\beta K} \ldots]\,
\Big|_{\beta =2\pi}\,,
\label{I2}
\ee
with the understanding that `everything else' is kept fixed
(c.f. below). As in the case of the conformal anomaly such a response 
has two aspects. One is to define the systems with $\beta \neq 2\pi$, --
which can no longer be ordinary Minkowski space QFTs. The other 
concerns the evaluation of the response itself, which again has 
significance within the context of Minkowski space QFT.

Let us now elucidate on the $\beta \neq 2\pi$ systems. There are 
two very different notions of ``varying the Unruh temperature'':
\begin{itemize}
\item[(a)] The classical notion of changing the norm of the 
timelike Killing vector field $\cK$ in $W$ (and hence the 
surface acceleration $a$ of the Rindler horizon).
\item[(b)] The ``replica'' understanding of Callan and Wilczek 
\cite{CW} to replace $e^{2\pi K}$ in (\ref{I1}) with 
$(e^{2\pi K})^{\beta/2\pi} = e^{\beta K}$,
keeping everything else (state space, operator products etc.) fixed.
\end{itemize}
There is a certain danger of confusing both notions because also the 
purely classical variation (a) affects the ($\hbar$-dependent) Unruh 
temperature $T= \frac{\hbar}{k c}\frac{a}{2\pi}$. To see this recall that 
in inertial coordinates $\cK = a(x^1\partial_0 + x^0\partial_1)$, and
its norm is $\|\cK\| = a \sqrt{- x^2}$, using metric conventions 
$ds^2 = (dx^0)^2 -(dx^1)^2$ and $W = \{x \in \R^{1,1} \,|\, 
|x^0| \leq x^1\}$. The result (\ref{I1}) holds in any dimension, for 
simplicity we specialize to 1+1 dimensions already at this point.
Clearly $\cK$ is unique up to normalization and changing its norm amounts to 
changing the Unruh temperature according to $T_1/T_2 = 
\|\cK_1\|/\|\cK_2\|$. In particular (a) leaves the classical spacetime 
intact and thus is not the relevant concept if one wants to compute 
a quantum response of the form (\ref{I2}) or unravel the statistical 
origin of the Horizon entropy \cite{GeoSa,GeoSb,CW,KSS,KS,HLW}. 
Henceforth we shall exclusively be concerned with 
the notion (b) of varying the Unruh temperature. In order to 
disentangle both aspects we fix the norm of $\cK$ once and for 
all to be $\|\cK\| = \sqrt{-x^2}$. For the generator $K$ of the 
Lorentz boosts in the Minkowski space QFT this means
\ba
&& e^{i \lb K}\cO(x) e^{-i\lb K} = \cO(x(\lb))\;,\;\;\;\mbox{where} 
\nonum
&&
x^0(\lb) = x^0 \ch \lb + x^1 \sh\lb\;,\;\;\;
x^1(\lb) = x^0 \sh\lb + x^1 \ch\lb\;.
\label{I3}
\ea
Note that with these normalizations a spacetime reflection 
corresponds to a complex Lorentz boost by $i\pi$.

Having fixed the normalization (\ref{I3}) the ``replica'' 
understanding of taking $\beta$ off $2\pi$ has the 
immediate consequence that one is no longer dealing with an 
ordinary QFT system. Taking again the heuristic formula (\ref{I1})
as a guideline one sees that translation invariance is broken:
$\Tr[e^{\beta K} U(x) \cO U(x)\inv] = \Tr[e^{\beta K} \cO U(x)\inv 
U(x(-i\beta))]$, if $U(x)$ is the unitary representation of the 
translation  group in the original QFT. Probably this should be
viewed as the Minkowski space version of the conical singularity
encountered in the Euclidean approach to the $\beta \neq 2\pi$ 
systems \cite{Dow,BTZ,Mor,MorV}.

\newsubsection{Thermalization without event horizons}

Here we shall pursue an approach to computing responses of the form
(\ref{I2}) which preserves the Lorentzian signature and which implements
the ``replica'' understanding of the $\beta \neq 2\pi$ systems on the 
level of form factors. It is based on a thermalization phenomenon related 
but not identical to (\ref{I1}) \cite{MNcycl}. 
A schematic comparison of both phenomena is given in the table below.
\vspace{5mm}

\hspace{1.7cm}
\begin{tabular}{c|c}
Unruh Thermalization & Form Factor Thermalization  \\[0.5ex] 
\hline 
Hyperbolic World lines in & Mass hyperboloids in \\
Rindler space             & forward light cone\\
$x^0 = r \cosh \tau,\;\;  x^1 = r \sinh\tau $ & 
$p_0 = m \cosh \th,\;\;p_1 = m \sinh\th $  \\
$\tau$: Killing time      & $\theta$: rapidity\\
\hline
Wightman functions in & Form Factors in \\
Rindler space obey $(KMS)_{2\pi}$ &
Minkowski space obey $(KMS)_{2\pi}$ \\
with respect to $\tau$ & with respect to $\th$      
\end{tabular}
\vspace{5mm}

$(KMS)_{2\pi}$ denotes a thermal equilibrium condition
of temperature $1/2\pi$ (with the normalization (\ref{I3}) 
and units $k = c = \hbar =1$). Mathematically, as first noticed by 
Kubo-Martin-Schwinger and Haag-Hugenholz-Winnink, the equilibrium 
condition gets encoded into certain analyticity properties with respect 
to some (generalized) time variable. The above form factor result has 
been shown to hold in {\em any} 1+1 dimensional QFT with a well-defined 
scattering theory, regardless of its integrability \cite{MNcycl}.
Form factors in this context are matrix elements of some field 
operator between the physical vacuum and the asymptotic multi-particle
states. Let us denote the $n$-particle form factor of a field 
operator $\cO$ symbolically by $\bra 0|\cO |A(p_n)\ldots A(p_1)\ket$, 
where $A(p)$ generates a 1-particle state of momentum 
$p= (m\cosh\th,m\sinh\th)$. Parallel to (\ref{I1}) one can give a 
mnemonic summary of the result as follows 
\be
\bra 0|\cO |A(p_n)\ldots A(p_1)\ket =
\Tr[e^{2\pi K} \cO \,A(p_n)\ldots A(p_1)]\;.
\label{ffcycl}
\ee
Here $K$ is again the generator of Lorentz boosts and the comments 
from footnote 1 apply likewise. These technical aspects aside, it may 
come as a surprise that matrix elements of a zero temperature QFT exhibit
thermal features, {\em without} an event horizon being invoked as 
in the Hawking-Unruh case. One explanation stems from the fact that 
the left hand side of (\ref{ffcycl}) does not make essential use of the 
micro-causality property of the QFT, while the thermal structure of the 
right hand side can be understood as a `remnant of micro-causality' on 
the level of scattering states. Another explanation can be gained 
from the strategy followed in the proof \cite{MNcycl}, which traces
the thermalization in (\ref{ffcycl}) back to that in the 
Bisognano-Wichmann-Unruh case \cite{BW, Unruh} by enclosing 
the `wave packets' eventually forming the scattering states 
into `comoving' wedge regions. In a forthcoming paper we study the 
generalization of (\ref{ffcycl}) to higher dimensions.

Using (\ref{ffcycl}) as a guideline it is easy to see what the 
replica understanding of $\beta \neq 2\pi$ amounts to. 
If we write $F(\th_n, \ldots, \th_1)$  for 
$\bra 0|\cO |A(p_n)\ldots A(p_1)\ket$  the defining properties 
should be 
\ba
&& F(\th_n+ i\beta,\th_{n-1}, \ldots ,\th_1) =
\eta\,F(\th_{n-1},\ldots,\th_1,\th_n)\;,\;\;\;\mbox{and}\nonum
&& F(\th_n,\ldots,\th_1)\;\;
\mbox{has kinematical poles at}\;\;\th_k - \th_j = \pm i\pi\,,
\,k\neq j\;. 
\label{I5}
\ea
The first equation is the $(KMS)_{\beta}$ condition where a generalized
statistics phase $\eta$ may appear. The second requirement stems from
`keeping everything else fixed'. In particular the normalization 
(\ref{I3}) should be kept fixed so that in the parameterization
$p= (m\cosh\th,m\sinh\th)$ a sign flip still corresponds to 
a complex Lorentz boost by $i\pi$ rather than $i\beta/2$. This entails that 
the position of the kinematical singularities in rapidity space stay
at $\th_k - \th_j = \pm i\pi$, $j\neq k$. Though (\ref{I5}) is a very 
natural transcription of the replica idea, it has a number of unexpected 
consequences. For example the spectra of the conserved charges on 
a multi-particle state change in a nontrivial way, c.f. section 3. 
The requirements (\ref{I5}) can be taken as the defining features for 
the $\beta \neq 2\pi$ systems in any 1+1 dim. QFT with a well-defined
scattering theory. Very likely also a generalization to higher
dimensions is possible. However the conditions are particularly 
stringent in so-called integrable QFTs on which we shall focus 
from now on.

By definition integrable massive QFTs are those for which the 
scattering operator enjoys a certain factorization property.
This allows one to express all S-matrix elements in terms of 
the two-particle (``bootstrap'') S-matrix, which in turn is a solution of the 
Yang-Baxter equation. For such QFTs the so-called form factor 
approach allows one to characterize the full non-perturbative dynamical
content of the QFT in terms of a recursive system of functional equations
known as ``form factor equations''. These functional equation only take 
the two-particle S-matrix as an input and are not 
renormalized or modified in any way in the process of solving the 
theory. Further they entail the quantum field theoretical locality 
requirement on the level of the Wightman functions \cite{Smir}. It 
thus seems natural to modify this system of functional equations 
such that (\ref{I5}) is obeyed. It turns out that this can be done
in a way that preserves the bootstrap S-matrix. For integrable QFTs 
therefore the concept of ``keeping everything else fixed'' in the 
replica understanding (b) of the $\beta \neq 2\pi$ systems acquires 
the precise meaning of: ``Keeping the bootstrap S-matrix fixed''. 
Indeed, since ultimately the entire QFT gets constructed from it, 
changing the bootstrap S-matrix would amount to changing the theory. 
Remarkably with this specification mathematical consistency then 
dictates almost everything else \cite{MNmod}. Most importantly the 
residue equations have to be modified in a certain way. The sequences 
of meromorphic functions solving the modified functional equations no 
longer define the form factors of a relativistic QFT. We shall see:
\begin{itemize}
\item Each integrable QFT admits a 1-parameter deformation 
that preserves the bootstrap S-matrix. The deformation parameter $\beta$ 
plays the role of an inverse temperature in the ``replica'' understanding
of taking $\beta$ off $2\pi$ in (\ref{ffcycl}). 
\item The form factor approach provides an explicit computational 
scheme for these systems, just as for $\beta =2\pi$.
It yields a finite, cutoff-independent answer for the response 
(\ref{I2}) of any local QFT quantity. 
\item For $\beta \neq 2\pi$ the underlying kinematical arena
is deformed, but is by construction compatible with the 
full non-perturbative dynamics of the system. Lorentz invariance is 
maintained exactly.
\end{itemize}
The rest of this paper is organized as follows. In the next section 
we briefly describe the modified form factor equations and 
show that their solutions can have a regular $\beta\ra 2\pi$ limit 
producing ordinary form factors. In section 3 we compute the spectrum of the
conserved charges in the deformed theory with some details relegated
to an appendix. Further the form factor resolution of the deformed 
two-point functions is introduced and its use illustrated for the 
energy momentum tensor of a free boson or fermion. In section 4
finally a few sample form factors in the deformed Ising model and 
Sinh-Gordon model are computed to demonstrate the feasibility of 
the scheme for interacting QFTs. The setting described bears some 
resemblance to 't Hoofts ``S-matrix Ansatz'' \cite{tHooftS,tHooft}. 
On this and other interrelations we briefly comment in the conclusions. 
\newpage 

\newsection{Deformed form factor equations}

Let $S_{ab}^{cd}(\th)$, $\th \in \C$, be a bootstrap S-matrix without 
bound state poles, i.e.~a matrix-valued meromorphic function analytic in
the strip $0 \leq {\rm Im}\,\th < \pi$ and satisfying the Yang-Baxter 
equation, unitarity and crossing. The indices $a,b,\ldots$ refer to a basis 
in some finite dimensional vector space $V$. Raising and lowering of 
indices is done by means of the charge conjugation matrix $C_{ab}$ 
and its inverse $C^{ab}$. To any such S-matrix one can associate a 
1-parameter family of functional equations, whose solutions are 
sequences tensor-valued meromorphic functions
\cite{MNmod}. The consistency of these equations is most conveniently
seen in an algebraic implementation. Here we shall give only a minimal
set of equations, which -- taken for granted the consistency of 
the deformation -- entail all others. For generic $\beta \neq 2\pi$ 
thus consider the following set of functional equations \cite{MNmod}
\begin{subeqnarray}
&\nspace& F_{a_n\ldots a_1}(\th_n+ i\beta,\th_{n-1},\ldots,\th_1) =
\eta\,F_{a_{n-1}\ldots a_1 a_n}(\th_{n-1},\ldots,\th_1,\th_n)\;,
\\
&\nspace&  F_{a_n\ldots a_1}(\th_n,\th_{n-1},\ldots ,\th_1) =
S_{a_na_{n-1}}^{d\;\;c}(\th_{n,n-1}) \,
F_{c\,d a_{n-2}\ldots a_1}(\th_{n-1},\th_n,\th_{n-2} \ldots ,\th_2,\th_1)\\ 
&\nspace& \mbox{Res}\,F^{(n)}_A(\th_{n-1} + i\pi,\th_{n-1},\th_{n-2},
\ldots,\th_1)
= - \lb^- C_{a_n a_{n-1}}\,F^{(n-2)}_{a_{n-2}\ldots a_1}
(\th_{n-2},\ldots,\th_1)\;.
\label{deff1}
\end{subeqnarray}
For $\th_n = \th_{n-1} -i\pi$ one has 
\addtocounter{equation}{-1}
\begin{subeqnarray}
\addtocounter{subequation}{3}
&\nspace& \mbox{Res}\,F^{(n)}_A(\th_{n-1} - i\pi,\th_{n-1},\th_{n-2},
\ldots,\th_1)
= - \lb^+ C_{a_n a_{n-1}}\,F^{(n-2)}_{a_{n-2}\ldots a_1}
(\th_{n-2},\ldots,\th_1)\;,\nonum
&\nspace &\mbox{Res}\,C^{a_na_{n-1}}
F^{(n)}_A(\th_{n-1}-i\pi,\th_{n-1},\th_{n-2},
\ldots,\th_1)
= - \lb^{-}\,F^{(n-2)}_{a_{n-2}\ldots a_1}(\th_{n-2},\ldots,\th_1)\;.
\label{deff1d}
\end{subeqnarray}
The first version applies to a $2\pi i$-periodic S-matrix, the second
when $S_{ab}^{cd}(-i\pi)$ is singular. 
Here and below ``Res'' denotes the residue at the simple pole of the 
displayed pair of rapidities, here: $\mbox{Res} = 
\mbox{res}_{\th_{n} = \th_{n-1} \pm i\pi}$. Further $A=(a_n,\ldots,a_1)$,
$\th = (\th_n,\ldots,\th_1)$ and we use the shorthand 
$\th_{kj} := \th_k - \th_j$ throughout. The constants 
$\lb^- = -i\beta/\pi,\;\lb^+ = \lb^-/\dim V$ are chosen to match the 
normalization of the 1-particle states ${}_b \bra \th_2|\th_1 \ket_a =
2\beta \delta_{ba}\delta(\th_{21})$, and $\eta$ is a complex phase.
In addition to the equations 
(\ref{deff1}.a-d) of course one has to specify the analytic structure of 
the solutions aimed at. For $\beta/2\pi$ irrational we require 
the solutions of (\ref{deff1}.a,b) to be meromorphic functions with poles 
at most at $\th_{kj} = \pm i\pi$ modulo $i\beta$; in particular they 
are supposed to be regular at $\th_{j+1,j} = i\beta$. The equations
(\ref{deff1}.c,d) then serve to arrange the solutions of (\ref{deff1}.a,b) 
into sequences. One aspect of the consistency alluded to before is that the 
operations `application of a symmetry transformation' via 
(\ref{deff1}.a,b) and `taking the residue' via (\ref{deff1}.c,d) commute. 
In particular this implies that any solution of (\ref{deff1}.a,b) having 
a simple pole at $\th_n = \th_{n-1} \pm i\pi$ will have simple poles also 
at $\th_{j+1,j} = \pm i\pi$ and $\th_{j+1,j} = \mp i(\pi -\beta)$, whose 
residues are given by 
\begin{subeqnarray}
&& \mbox{Res}\,
F^{(n)}_A(\th_n,\ldots,\th_{j} + i\pi,\th_j,\ldots,\th_1)\nonum
&&\sspace = - \lb^- C_{a_{j+1}a_j}\,
F^{(n-2)}_{a_n\ldots a_{j+2}a_{j-1}\ldots a_1}
(\th_n,\ldots,\th_{j+2},\th_{j-1},\ldots ,\th_1)\;,\\
&& \mbox{Res}\,F^{(n)}_A(\th_n,\ldots,
\th_j-i\pi +i\beta,\th_j,\ldots,\th_1) \nonum
&&\sspace 
= -\lb^+\,
L_{t_{j+1}}(\th_n,\ldots,\th_j-i\pi +i\beta,\th_j,\ldots,\th_1)_A^B\;
C_{b_{j+1}b_j}\,F^{(n-2)}_{p_jB}(p_j\th)\;,
\label{deff2}
\end{subeqnarray}
and similar equations for  $\th_{j+1,j} = -i\pi$ and 
$\th_{j+1,j} = i(\pi -\beta)$. The notation is
$p_jA=(a_n,\dots a_{j+2}a_{j-1}\ldots a_1)$ and 
$p_j\th =(\th_n,\ldots,\th_{j+2},\th_{j-1},\ldots,\th_1)$. Further
\be
L_{\tau_j}(\th)_A^B =  \eta\,T_{a_j}^c(\th_j|\th_n,\ldots,\th_{j+1})%
^{b_n\ldots b_{j+1}}_{a_n\ldots a_{j+1}}\;
T_c^{b_j}(\th_j-i\beta|\th_{j-1},\ldots,\th_1)%
_{a_{j-1}\ldots a_1}^{b_{j-1}\ldots b_1}\;,
\label{deff3}
\ee 
is the matrix entering the deformed Knizhnik-Zamolodchikov equations; 
the corresponding action on rapidity vectors is $\tau_j(\th_n,\ldots,\th_1) =
(\th_n,\ldots,\th_j + i\beta, \ldots, \th_1)$. As indicated it can be 
expressed in terms of the monodromy matrix  
$$
T_{a_n}^{b_n}
(\th_n|\th_{n-1},\ldots,\th_1)_{a_{n-1}\ldots a_1}^{b_{n-1}\ldots b_1} 
= S_{c_{n-1}a_{n-1}}^{b_n b_{n-1}}(\th_{n-1,n})\;
S_{c_{n-1}a_{n-2}}^{c_{n-2} b_{n-2}}(\th_{n-2,n})\;\ldots\;
S_{a_na_1}^{c_2b_1}(\th_{1,n})\;,
$$
whose trace over $a_n=b_n$ yields the
well-known family of commuting operators on $V^{\otimes (n-1)}$.

The dependence on $\beta$ in the deformed form factors will usually be 
suppressed. When needed to distinguish them from the undeformed form
factors we shall write $(F^{(\beta,n)})_{n\geq 0}$ and 
$(F^{(2\pi,n)})_{n\geq 0}$ for the deformed and undeformed ones, respectively. 
In this notation one can select solutions for generic $\beta$ such that 
\be
F^{(2\pi,n)}_A(\th) =\lim_{\beta \ra 2\pi} F_A^{(\beta,n)}(\th)\;.
\label{deff4}
\ee
To verify this one has to show that the right hand side solves the 
undeformed form factor equations. For the equations (\ref{deff1}.a,b) 
this is obvious. To see that the residue equations come out correctly 
observe that in the limit $\beta \ra 2\pi$ the poles at 
$\th_{j+1,j}= i\pi$ and 
$\th_{j+1,j}= i(\beta -\pi)$ in (\ref{deff2}) merge. They produce 
a simple pole again because by assumption $F^{(\beta,n)}_A(\th)$ does 
not have a pole at $\th_{j+1,j}= i\beta$. In particular this implies 
that the residues of the merged poles add up producing 
\ba
&\nspace &\mbox{Res}\,F^{(n)}_A(\th_n,\ldots,
\th_j+i\pi,\th_j,\ldots,\th_1) \nonum
&\nspace &\sspace 
=- \left[ \lb^+ L_{\tau_{j+1}}%
(\th_n,\ldots,\th_j+i\pi,\th_j,\ldots,\th_1)_A^B\;
+ \lb^- \delta_{A}^{B}\right]\,C_{b_{j+1}b_j} F^{(n-2)}_{p_jB}(p_j\th)\;,
\label{deff5}
\ea
which is the undeformed residue equation. In principle it is non-trivial
that solutions of the deformed equations exist such that (\ref{deff4})
is satisfied. Based on experience with the simple models described later,
we expect however the following to be true:
\begin{itemize}
\item For each ordinary form factor sequence 
$(F^{(2 \pi, n)})_{n \geq 1}$ there {\em exists} a deformed counterpart
$(F^{(\beta, n)})_{n \geq 1}$ such that (\ref{deff4}) is satisfied. 
\item The deformed sequence is in general not uniquely specified
by (\ref{deff4}) but can be made so by imposing suitable 
minimality conditions.
\end{itemize}
Naturally one will search for solutions of the deformed equations
with a definite degree of homogeneity $s$ (``spin'') under 
the action of $e^{i\lb K}$, $iK = \sum_{j} \frac{\partial}{\partial\th_j}$. 
In the undeformed case suitable multiplets of solutions then transform
according to tensor representations of SO(1,1), reflecting the Lorentz
covariance properties of the local operator assigned to it. The 
same can be done here, though a-priori without any reference to an 
underlying QFT system. For example an appropriate triplet of 
solutions $F_s^{(n)}(\th)$ of spin $s = 0,\pm 2$ can be used to define a 
symmetric second rank SO(1,1) tensor $F_{\mu \nu}^{(n)}(\th)$,
\ba
&& F^{(n)}_{\mu \nu}(\th_n + \lb,\ldots, \th_1 + \lb) = 
\Lambda(\lb)_{\mu}^{\;\rho}
\Lambda(\lb)_{\nu}^{\;\sigma}\, F^{(n)}_{\rho \sigma}(\th)\;,\nonum
&& \Lambda(\lb)_{\mu}^{\;\nu} = 
\left(\begin{array}{cc}
\ch\frac{2\pi}{\beta}\lb &\sh\frac{2\pi}{\beta}\lb\\ 
\sh\frac{2\pi}{\beta}\lb &\ch\frac{2\pi}{\beta}\lb 
\end{array}\right)\;,\;\;\;\mu,\nu = 0,1\;,
\label{deff6}
\ea
where the components $F_{\mu\nu}^{(n)}(\th)$ are linear combinations of 
$F_s^{(n)}(\th)$, $s = 0,\pm 2$.

\newsection{Deformed kinematics}
  
The structure of the deformed kinematics turns out to be largely 
dictated by consistency with the dynamics, i.e.~with the deformed 
form factors equations. In this section we present some aspects of the 
resulting kinematics.

\newsubsection{Deformed conserved charges} 

In the undeformed case local conserved charges are characterized 
by two properties. They act 
numerically on asymptotic multi-particle states and their eigenvalues 
decompose into a sum of 1-particle contributions. On the level of
form factors, the first property implies that the eigenvalues are 
trivial solutions of the form factor equations. Here we take the 
(deformed) form factor equations as the starting point, so that
it is natural to define a conserved charge in terms of its
eigenvalues as follows: A conserved charge $Q_s$ of spin $s$ has 
eigenvalues $Q_s^{(n)}(\th)$ that are real for real arguments, 
$i\beta$-periodic and symmetric in all variables, as well as homogeneous 
and hermitian in the following sense
\be
Q_s^{(n)}(\th_n +\lb,\ldots,\th_1 +\lb) = e^{s \frac{2\pi}{\beta}\lb}\,
Q_s^{(n)}(\th)\;,\sspace Q_s^{(n)}(\th)^* = Q_s^{(n)}(\th^*)\;.
\label{charge2}
\ee
Further the eigenvalues for $n$ and $n-2$ particles are linked by 
the recursive relation
\be
Q_s^{(n)}(\th_n=\th_{n-1}\pm i\pi,\th_{n-1},\th_{n-2},\ldots,\th_1) =
Q_s^{(n-2)}(\th_{n-2},\ldots,\th_1)\;.
\label{charge3}
\ee 
In summary a conserved charge in the deformed theory is
in correspondence to a sequence $(Q_s^{(n)}(\th))_{n \geq 0}$ of 
symmetric functions solving (\ref{charge2}), (\ref{charge3}). 
The structure of the solutions turns out to be quite different
from that for $\beta =2 \pi$. Intrinsically however the role of 
the deformed eigenvalue sequences is precisely the 
same as in the undeformed case: Pointwise multiplication of a given 
form factor sequence with $(Q_s^{(n)}(\th))_{n \geq 0}$ produces a
new form factor sequence (with spin $s + s'$, if the original sequence
had spin $s'$). Clearly the set of eigenvalue sequences forms a 
graded abelian ring with respect to pointwise addition and multiplication, 
where upon multiplication the degrees add up. In a theory whose 
S-matrix has bound state poles
the recursion relation (\ref{charge3}) will be supplemented by a $n \ra n-1$
recursive relation, which in particular serves as a selection principle 
for the allowed spin values. Here we shall restrict attention to S-matrices
without bound state poles. In particular the mass gap $m$ then provides
the only intrinsic mass scale of the theory.

For $\beta = 2\pi$ the most important solutions of (\ref{charge2}), 
(\ref{charge3}) are the ``power sums'' 
$P_s^{(n)}(\th) \sim t_1^s + \ldots + t_n^s$ for $s$ odd and 
$t_j = e^{\th_j}$. In fact these powers sums form a basis for the 
before-mentioned ``ring of conserved charges'' at $\beta =2\pi$ \cite{Ising5}.
That is to say all other solutions of  (\ref{charge2}), 
(\ref{charge3}) are linear combinations of products of the power sums.
In physical terms $P_{\pm 1}^{(n)}(\th)$ for example 
are (up to a normalization constant with units of a mass) the eigenvalues 
of the lightcone momenta $P_{\pm}$ on an asymptotic $n$-particle state. 
Their product $P_{+1}^{(n)}(\th)P_{-1}^{(n)}(\th)$ is proportional 
to the $n$-particle eigenvalues of the $(\rm{mass})^2$ operator 
$2 P_+ P_-$. Observe also that the power sums (and 
their linear combinations) are distinguished by the property that 
for them the rapidities provide an additive parameterization of the 
multi-particle eigenvalues
\be
P_s^{(n)}(\th) = \sum_j P_s^{(1)}(\th_j)\;.
\label{charge4}
\ee
For $\beta \neq 2\pi$ such solutions  of (\ref{charge2}), 
(\ref{charge3}) no longer exist. Nevertheless natural deformed 
counterparts of the power sums do exist and they will play a key role
in the following.

The starting point for the construction of the deformed power sums 
is the following fact: Let $P^{(n)}(N,l)$ be the space 
of symmetric polynomials in $t_1,\ldots ,t_n$ of total degree $N$ and 
partial degree $l$ (c.f. Appendix). There exists a unique 
sequence of $i\beta$-periodic symmetric functions 
$(P^{(n)}(t))_{n \geq 0}$ such that 
\begin{itemize}
\item[(a)] $P^{(n)}(t)$ is a ratio of symmetric polynomials in
$t_j = e^{2\pi \th_j/\beta}$, $1 \leq j\leq n$, solving 
(\ref{charge2}), (\ref{charge3}). The numerator $\num P^{(n)}$ and 
denominator $\den P^{(n)}$ have the following degrees
\be
\num P^{(n)} \in P^{(n)}(N+1, n)\;,\;\;\;   
\den P^{(n)} \in P^{(n)}(N, n-1)\;,\;\;\; N= \frac{1}{2}n(n-1)\;.
\label{charge5}
\ee
\item[(b)] $P^{(n)}(t)$ is proportional to the eigenvalue of the 
$P_+$ lightcone momentum operator for $\beta \ra 2\pi$, i.e.
\be
P^{(n)}(t)\bigg|_{\beta = 2\pi}= t_1 + \ldots + t_n\;.
\label{charge6}
\ee
\end{itemize}
Further no solution of the same type with lower degrees in 
(\ref{charge5}) exists. Here and later on it will be convenient to 
express all symmetric polynomials in terms of the elementary symmetric 
polynomials $\s_k^{(n)}$, $k =1,\ldots, n$ (where we shall usually suppress 
the superscripts $n$). Writing further $\g =2\cos\frac{\pi^2}{\beta}$ 
and $\d = 2\cos\frac{2\pi^2}{\beta} = q + q\inv = \g^2 -2$, the 
explicit results for $n \leq 4$ are:
\ba
&& \num P^{(2)} = \s_1^2 - \g^2 \s_2\;, 
\sspace  \den P^{(2)} = \s_1\;,
\nonum
&& \num P^{(3)} = \s_1^2\s_2 - \g^2\s_2^2 + 
(1+\d)\s_1\s_3 \;,
\nonum
&& \den P^{(3)} = \s_1\s_2 - (1 + \d)^2\,\s_3\;,
\nonum
&& \num P^{(4)} = \s_1^2\s_2\s_3 - \g^2\,\s_2^2\s_3 + 
(1 + \d)\s_1\s_3^2 -(1 + \d)^2 \s_1^3 \s_4 
\nonum 
&& \bspace + \d\g^4\,\s_1\s_2\s_4 - \d^2 \g^2(1 + \d)\,\s_3\s_4\;,
\nonum
&& \den P^{(4)} = \s_1\s_2\s_3 - (1 + \d)^2(\s_3^2 + \s_1^2 \s_4) 
+ \d \g^4\,\s_2\s_4\;.
\label{charge7}
\ea
The most efficient way to compute these expressions is by making use 
of the fact that both the numerators and the denominators separately obey a 
recursive relation. These and other aspects of the construction of the 
conserved charge eigenvalues are relegated to the Appendix. Clearly the 
solutions (\ref{charge7}) violate (\ref{charge4}). Nevertheless one can 
recover an additive parameterization by means of the following result. 
\medskip

{\bf Lemma:} Let $P^{(n)}(t)$ be the functions defined above. Set 
\be
l_j^{(n)}(t) := t_j\dd{t_j} P^{(n)}(t)\;,\;\;\;j=1,\ldots,n\;,
\label{l1}
\ee
which by construction reduce to $t_j$ for $\beta = 2\pi$. Then
\be
P_s^{(n)}(t) := 
[l_1^{(n)}(t)]^s + \ldots + [l_n^{(n)}(t)]^s\;,
\label{l2}
\ee
is a conserved charge eigenvalue for all (positive and negative) odd 
integers $s$, which for $\beta =2\pi$ reduces to $t_1^s +\ldots + t_n^s$.
\medskip

The point here is that the expressions (\ref{l2}) again
solve the recursive equation (\ref{charge3}). For $s =1$ this is just 
a rewriting of the Euler relation, but for $s \neq 1$ the proof is 
more involved. We omit it. Clearly the $P_s^{(n)}(t)$ are natural 
deformed counterparts of the power sums. 
They are again expressible as ratios of homogeneous symmetric 
polynomials, though usually of a fairly high (total and partial) degree. 

Let us focus now on the $s = \pm 1$ conserved charges. A drawback of
the construction (\ref{l1}), (\ref{l2}) is that the $s=1$ and the 
$s =-1$ charges enter asymmetrically. To understand how this comes about 
consider $P^{(n)}(t\inv)$ with $t\inv = (t_n\inv,\ldots,
t_1\inv)$, which meets the same requirements as $P_{-1}^{(n)}(t)$: 
It solves (\ref{charge2}), (\ref{charge3}) with $s=-1$ and reduces
to $t_n\inv + \ldots + t_1\inv$ for $\beta = 2\pi$. 
In terms of the elementary symmetric polynomials the inversion $t \ra t\inv$ 
amounts to the replacement $\s_k \ra \s_{n-k}/\s_n$. The 
functions $P^{(n)}(t\inv)$ are thus again expressible as ratios of 
symmetric polynomials in $t_1, \ldots, t_n$ with degrees readily 
worked out from (\ref{charge5}). If we next consider the sequence of 
ratios 
\be
Q^{(n)}(t) := \frac{P_{-1}^{(n)}(t)}{P^{(n)}(t\inv)}\;,\sspace
Q^{(n)}(t)\bigg|_{\beta = 2\pi} = 1\;,\;\;n \geq 1\;,
\label{momdef1}
\ee
its members qualify as spin zero solutions of (\ref{charge2}), 
(\ref{charge3}) starting with $Q^{(0)} = 1/(4 - \g^2)$ and 
$Q^{(1)} =1$. In other words the $s=1$ and $s =-1$ power sums in 
(\ref{l2}) feature asymmetrically only because one of them has 
been multiplied with a complicated spin zero conserved charge
having trivial $\beta = 2\pi$ limit.  
It is obviously nicer to distribute the square root of $Q^{(n)}(t)$ 
symmetrically on both the $s=1$ and the $s=-1$ power sums.
This leads us to the following definition of the 
deformed lightcone momentum eigenvalues
\ba
&& k_j^{(n)}(t) := [Q^{(n)}(t)]^{1/2} \,l_j^{(n)}(t)\;,\;\;\;
j =1,\ldots, n\;,\nonum
&& P_+^{(n)}(t) := \frac{m}{\sqrt{2}} \sum_{j=1}^n k_j^{(n)}(t) =
\frac{m}{\sqrt{2}} [Q^{(n)}(t)]^{1/2} \,P^{(n)}(t)\;,\nonum
&& P_-^{(n)}(t) := \frac{m}{\sqrt{2}} \sum_{j=1}^n [k_j^{(n)}(t)]^{-1} =
\frac{m}{\sqrt{2}} [Q^{(n)}(t)]^{1/2} \,P^{(n)}(t\inv)\;,
\label{momdef2}
\ea
where $m$ is the mass gap. In particular in this way an additive 
parameterization and a standard relativistic dispersion 
relation are recovered. Of course one would like to 
interpret $k_j^{(n)}(t)$ as the lightcone momentum of the $j$-th particle
in an $n$-particle state of the deformed theory. For this to be 
possible the $k_j^{(n)}(t)$ should better be non-negative functions
on $\R_+^n$. For sufficiently small $\g$ one expects this to work out,
but it is not obvious how large $\g$ can be made without sacrificing
this property. From the explicit expressions we verified 
that for $n \leq 4$  
\be
k_j^{(n)}(t) \geq 0\;,\;\;\forall t \in \R_+^n\;\;\;
\mbox{if}\;\;\g^2 <1\,.
\label{momdef3}
\ee 
In fact it is sufficient to check (\ref{momdef3}) for the $l_j^{(n)}(t)$,
because then also $Q^{(n)}(t)$ is nonnegative for $\g^2 <1$. 
We expect (\ref{momdef3}) to be a generic feature and henceforth restrict 
attention to $-1 <\g <1$, i.e. to $\frac{3}{4} < |\frac{\beta}{2\pi}| <
\frac{3}{2}$. 

\begin{figure}[htb]
\vskip 25mm
\leavevmode
\hskip 30mm
\epsfxsize=80mm
\epsfysize=60mm
\epsfbox{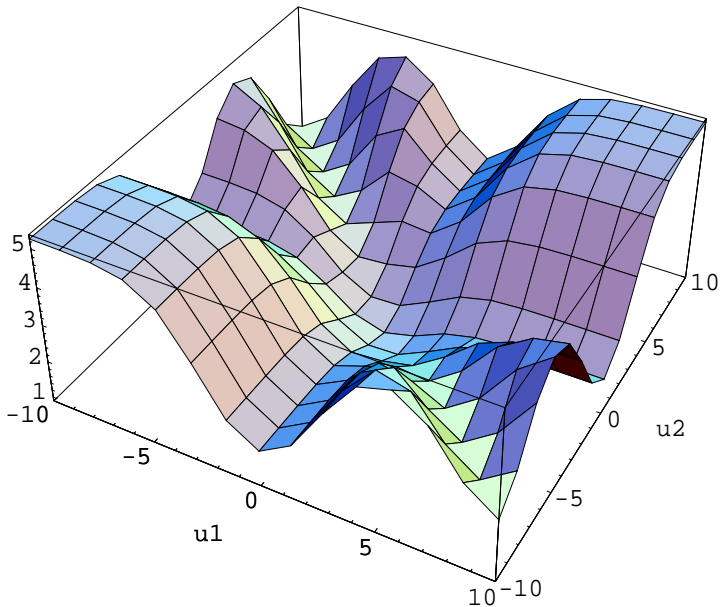}
\vskip 2mm
\caption{Ratio of the deformed and undeformed three-particle mass 
eigenvalues $M^{(3)}(u_1,u_2)/[M^{(3)}(u_1,u_2)]_{\beta = 2\pi}$ 
for $\g = 2 \cos \pi^2/\beta = 0.9$.}
\label{mass3}
\end{figure}

A natural definition of the deformed mass eigenvalues is 
\be
M^{(n)}(t) = [2 P_+^{(n)}(t) P_-^{(n)}(t)]^{1/2}\;.
\label{momdef4}
\ee
They are conserved charges with the correct $\beta \ra 2\pi$ limit.
In addition their threshold values are the same as in the undeformed 
case, i.e. $M^{(n)}(t) \geq n m$, $\forall t \in \R_+^n$, and equality 
only holds on the main diagonal of $\R^n_+$. Off the diagonal however the 
deformed mass eigenvalues are always larger than the undeformed ones. 
If we regard (\ref{momdef4}) as a function $M^{(n)}(u)$ of the 
rapidity differences $u_i = \frac{2\pi}{\beta}(\th_{i+1} - \th_i)$, 
$i=1,\ldots ,n-1$, this amounts to 
\be
M^{(n)}(u) \geq 
m\left[n + 2 \sum_{i < j}\ch(u_i + \ldots + u_j)\right]^{1/2} 
= M^{(n)}(u)\bigg|_{\beta = 2\pi}\;,
\label{momdef5}
\ee
where for $u_i \neq 0$ strict inequality holds. In physical terms 
(\ref{momdef5}) means that boosting two particles relative to each other 
costs more energy than in the undeformed case. The price can be measured 
in units of the undeformed energy, i.e.~in terms of the ratio
$M^{(n)}(u)/[M^{(n)}(u)]_{\beta = 2\pi}$. The resulting cost functions 
have a global minimum at $u =0$ and local minima in the 
form of `valleys' in the vicinity of the diagonals where two or more
rapidities coincide. Off the diagonals the ratio quickly approaches
a constant value. The height $h$ of this plateau rapidly increases
with $n$ and $|\g|$. For example at $\g = 0.9$ one has 
$h \approx 2.3,\, 5.2,\, 12$, for $n=2,3,4$, respectively. 
For $n =3$ and $\g =0.9$ the surface of the relative energy costs
is shown in Figure \ref{mass3}. For other values of $0<|\g|<1$ the 
surfaces are qualitatively similar.

\newsubsection{Rapidity diffeomorphisms}

\begin{figure}[htb]
\vskip 5mm
\leavevmode
\hskip 30mm
\epsfxsize=110mm
\epsfysize=55mm
\epsfbox{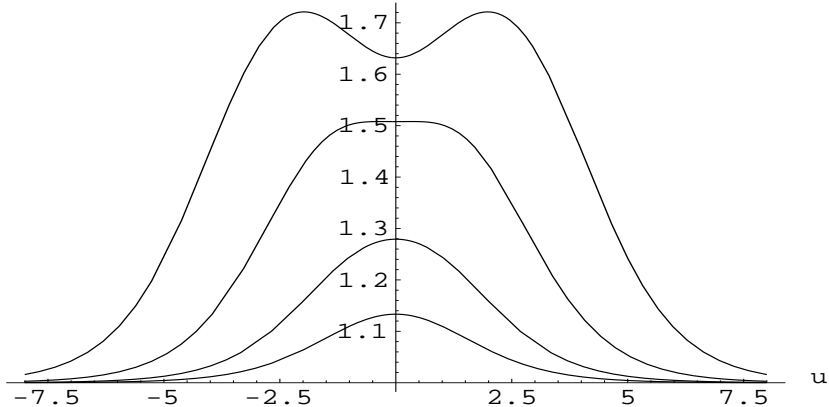}
\caption{Jacobian of the $n=2$ rapidity diffeomorphism $\Omega^{(2)}(u)$
for various values of $\g =2\cos \pi^2/\beta$. In order of increasing
maxima $\g = 0.5,\, 0.7,\, 0.9,\, 0.98$.}
\label{Omega}
\end{figure}

The positivity (\ref{momdef3}) also allows one to introduce single
valued `physical' rapidities $\alpha_j$ carrying an 
induced action of Lorentz boosts 
\be
\alpha_j^{(n)}(\th) := \frac{\beta}{2\pi} \ln k_j^{(n)}(t)\;,
\;\;\;\alpha_j^{(n)}(\th_n +\lb,\ldots, \th_1 + \lb) = \lb + 
\alpha_j^{(n)}(\th)\;.
\label{diff1}
\ee
The mapping $\R^{(n)}_+ \ra \R^{(n)}_+,
(t_n,\ldots, t_1) \ra (k_n,\ldots,k_1)$ is 
obviously differentiable and the Hessian can be checked to 
be nonsingular. Thus geometrically (\ref{diff1}) provides a 
diffeomorphism 
\be
\R^n \rra \R^n,\;\;\; (\th_n, \ldots, \th_1) \rra
(\alpha_n,\ldots ,\alpha_1)\;,
\label{diff2}
\ee
from the real `form factor' rapidities $\th_j$ to the real `physical' 
rapidities $\alpha_j$. The former have the virtue that in terms of them 
form factors,
conserved charge eigenvalues etc. admit an analytic continuation with 
controllable analyticity properties, which moreover adhere to the 
``replica'' understanding of taking $\beta \neq 2\pi$. However they
do not provide an additive parameterization of energy and momentum
on a multi-particle state. The latter can be achieved by switching to the 
rapidities $\alpha_j$ at the expense of a vastly more complicated 
structure of nontrivial form factors. (For example the form factors
constructed in section 4 re-expressed in terms of the $k_j$'s 
would be horrendous). Technically it is therefore convenient to 
work with the `form factor' rapidities $\th_j$ throughout taking into
account the Jacobian stemming from (\ref{diff2}). For the 
resolution of the identity in terms of multi-particle states this 
means    
\ba
\1 \is \sum_n \frac{1}{n!}\int 
\frac{d^n \alpha}{(2\beta)^n}|\alpha_n,\ldots,\alpha_1\ket
\bra \alpha_1,\ldots,\alpha_n| \nonum 
\is \sum_n \frac{1}{n!}\int 
\frac{d^n \th}{(2\beta)^n} \,\Omega^{(n)}(\th)
|\th_n,\ldots,\th_1\ket \bra \th_1,\ldots,\th_n|\;.
\label{diff3}
\ea
Since the form factors considered later will be functions of 
$t_j = e^{2\pi \th_j/\beta}$ only it is convenient to also 
treat the Jacobian as a function $\Omega^{(n)}(t)$ of the $t_j$'s.
This gives
\be
\int \frac{d^n \alpha}{(2\beta)^n} = 
\int \frac{d^n k}{(4 \pi)^n}\frac{1}{k_1 \cdots k_n} =
\int \frac{d^n t}{(4 \pi)^n}\frac{\Omega^{(n)}(t)}{t_1 \cdots t_n}=
\int \frac{d^n \th}{(2\beta)^n}\, \Omega^{(n)}(\th)\;, 
\label{diff4}
\ee
where the integrals are over $\R^n$ or $\R_+^n$ and  
\be
\Omega^{(n)}(t) = \frac{t_1 \ldots t_n}{k_1^{(n)} \ldots k_n^{(n)}}
\;\det \left( \frac{\partial k^{(n)}_i}{\partial t_j} 
\right)_{1 \leq i,j \leq n}\;.
\label{diff5}
\ee
As indicated we write $\Omega^{(n)}(\th)$ for $\Omega^{(n)}(t)$ when 
regarding the Jacobian as a function of the rapidities rather than their
exponentials. The measure $\Omega^{(n)}(t)$ is easily seen to have the 
following properties. It is again a ratio of symmetric polynomials and 
homogeneous of total degree zero. Its coefficients depend on $\beta$ 
only through $\gamma^2= (2\cos \pi^2/\beta)^2$ and it is positive for 
$\g^2 <1$. Except for $n=2$ reflection invariance is lost
$\Omega^{(n)}(t) \neq \Omega^{(n)}(t\inv)$. The explicit expressions
are in principle readily worked out. For example
\be
\Omega^{(2)}(t) = \frac{(1-\g^2)\s_1^4 + 4 \g^2 \s_1^2 \s_2 -\g^4 \s_2^2}%
{(1-\g^2)\s_1^4 + 2 \g^2 \s_1^2 \s_2 + \g^4 \s_2^2}\;.
\label{diff6}
\ee
Viewed as a function of the rapidities (\ref{diff6}) only depends on 
the difference $u = \frac{2\pi}{\beta}(\th_2 -\th_1)$. This function 
$\Omega^{(2)}(u)$ is displayed in Fig.\ref{Omega} below for various 
values of $\g$.  

This concludes our discussion of the momentum space kinematics.
We have computed $P_{\pm}^{(n)}(t)$ and $\Omega^{(n)}(t)$ 
explicitly in terms of elementary symmetric polynomials for $n \leq 4$.
The expressions are too long to be communicated
in print; however the files can be obtained from the author upon request.  
From $P_{\pm}^{(n)}(t), \Omega^{(n)}(t)$ and the list (\ref{charge7}) 
all other kinematical quantities considered can readily be obtained 
in explicit form.

\newsubsection{Deformed two-point functions}

Let us first consider the spacetime evolution of form factors.
Implicitly form factors refer to a fixed reference point in 
spacetime, which we have so far taken to be the `origin' of the wedge $W$. 
The evolution through spacetime simply amounts to multiplying with 
a phase factor $\exp[-i x \cdot P^{(n)}(\th)]$. In the deformed case 
we make a similar Ansatz%
\footnote{For simplicity we suppress internal indices here and later on.} 
\be
F^{(n)}(\th) \stackrel{x}{\rra} U_x^{(n)}(\th) F^{(n)}(\th)\;,
\label{Evol1}
\ee
where the assignment $\R^{1,1}\ni x \ra U_x^{(n)}(\th)$ has to 
obey various consistency conditions.
%
%
As a function of the rapidities $U_x^{(n)}(\th)$ must basically 
qualify as a conserved charge, just with modified homogeneity 
and hermiticity requirements. We thus take $U_x^{(n)}(\th)$
to be completely symmetric and $i\beta$-periodic in all the rapidity
arguments. Further it has to obey
\begin{subeqnarray}
&& U_x^{(n)}(\th_{n-1}\pm i\pi, \th_{n-1}, \th_{n-2},\ldots ,\th_1)
=U_x^{(n-2)}(\th_{n-2},\ldots ,\th_1)\;,\\
&& U_{\Lambda(\lb,x)}^{(n)}(\th_n +\lb,\ldots, \th_1 +\lb) =
U_x^{(n)}(\th)\;,\\
&&U_x^{(n)}(\th)^* = U_x^{(n)}(\th^*+i\pi)\;,
\label{Evol2}
\end{subeqnarray}  
where $\R \ni \lb \ra \Lambda(\lb,x) \in \R^{1,1}$ is some representation
of the boosts on $\R^{1,1}$. The first condition ensures consistency 
with the deformed residue equation, the second expresses boost invariance, 
and the third one is required by hermiticity and ``crossing'' 
(c.f. \cite{MNmod} for more details). The conditions (\ref{Evol2}) also
guarantee that the generalized form factors $\Fbar^{(m|n)}(\omega|\th)$,
$\omega =(\omega_m,\ldots,\omega_1)$, $\th =(\th_n,\ldots,\th_1)$ with
$m,n \geq 0$, evolve consistently. The latter are distributional kernels 
associated with a set of form factors $F^{(k)}$, $k = m + n -2l$, 
$l = 0,\ldots, \min(m,n)$. The explicit expression can be found in 
\cite{MNmod}, appendix A. The point relevant here is that, although
form factors with different particle numbers $k$ are involved, the 
condition (\ref{Evol2}a) allows one to combine the various terms to
obtain
\be
\Fbar^{(m|n)}(\omega|\th) \stackrel{x}{\rra} 
U^{(m+n)}_x(\omega + i\pi,\th) \,\Fbar^{(m|n)}(\omega|\th)\;.
\label{Evol3}
\ee
The property (\ref{Evol2}c) then also ensures that (\ref{Evol3}) is 
compatible with hermiticity and ``crossing'' of the generalized 
form factors, e.g. $[\Fbar^{(m|n)}(\omega^T|\th)]^* = 
\Fbar^{(n|m)}({\th^*}^T|\omega^*)$, if the original form factors 
are hermitian. In contrast to the undeformed case however
\be
U_x^{(m+n)}(\omega,\th) \neq U_x^{(m)}(\omega)\,
U_x^{(n)}(\th)\;.
\label{Evol4}
\ee
This means that $\Fbar^{(m|n)}(\omega|\th)$ cannot be 
interpreted as a matrix element $\bra\omega|\cO|\th\ket$ with an 
autonomous dynamics of the ``bra'' and ``ket'' vectors separately.  

It remains to find a solution of (\ref{Evol2}) that reduces to 
$\exp[-i x \cdot P^{(n)}(\th)]$ for $\beta \ra 2\pi$. The simplest 
solution is
\be
U_x^{(n)}(\th) = \exp\left\{ q^{-1/2} x^+ P_+^{(n)}(\th) 
- q^{1/2} x^- P_-^{(n)}(\th) \right\}\;,
\label{Evol5}
\ee
where $P^{(n)}_{\pm}(\th)$ are the deformed momentum eigenvalues 
(\ref{charge7}). The sum and difference of $P_{\pm}^{(n)}(\th)$
transforms according to the vector representation of SO(1,1), so that 
$\Lambda(\lb,x)$ in (\ref{Evol2}c) can likewise be taken to be the 
standard action 
of Lorentz boosts on $\R^{1,1}$. We can anticipate from (\ref{Evol5})
that ordinary translation invariance in the labels $x^{\pm}$ will 
be broken because $U_x(\th)^* \neq U_x^{(n)}(\th^*)\inv$.  
As noted in the introduction this is to be expected and can 
be regarded as a Lorentzian counterpart of the conical spaces
employed in the Euclidean approach to the $\beta \neq 2\pi$ systems
\cite{Dow,BTZ,Mor,MorV}. 

With these preparations at hand we can eventually introduce 
the form factor resolution for a deformed two-point function  
\be
W(x,y) = \sum_{n \geq 1} \frac{1}{n!} \int
\frac{d^n \th}{(2\beta)^n}\,\Omega^{(n)}(\th)
U_x^{(n)}(\th) U_y^{(n)}(\th)^*\,|F^{(n)}(\th)|^2\;.
\label{Evol9}
\ee
Despite the similarity to the undeformed case neither 
translation invariance nor micro-causality in the labels $x,y$ can 
be expected to hold for $\beta \neq 2\pi$. It should be interesting 
to see whether in an appropriate `quantum spacetime' these notions
can partially be restored. An examination of these issues is 
beyond the scope of the present paper. However for the computation of 
the response under a variation of $\beta$  
the information gathered so far is sufficient. In particular it is 
convenient to rewrite (\ref{Evol9}) in the form of a K\"{a}llen-Lehmann 
spectral representation. For simplicity let us assume that $F^{(n)}(\th)$ 
is of the form $F^{(n)}(\th) = [P_+^{(n)}(\th)]^{l_+}\,
[P_-^{(n)}(\th)]^{l_-}\,f^{(n)}(\th)$ with $l_{\pm}$ non-negative 
integers and $f^{(n)}(\th)$ a function of the rapidity differences 
only. Performing a change of integration variables
\be
u_j = \frac{2\pi}{\beta}(\th_j -\th_{j+1})\;,\;\;j =1,\ldots, n-1\;,\;\;\;
\alpha = \frac{\beta}{4\pi}\ln\frac{P_+^{(n)}(\th)}{P_-^{(n)}(\th)}
\ee
one finds 
\ba
&& W(x,y) = -i \sum_{n \geq 1} \int_0^{\infty} d\mu\, \rho^{(n)}(\mu)
\,(i\partial_{z^+})^{2l_+}(i\partial_{z^-})^{2l_-}D(z;\mu)\;,\nonum
&& \rho^{(n)}(\mu) = \int_0^{\infty}
\frac{du_1\ldots du_{n-1}}{(4\pi)^{n-1}}\, \Omega^{(n)}(u)
|f^{(n)}(u)|^2 \d(\mu - M^{(n)}(u))\;,\;\;n \geq 2\;,
\label{Evol10}
\ea
and $\rho^{(1)}(\mu) = \delta(\mu - m)|F^{(1)}|^2$. The deformed mass 
eigenvalues are given in (\ref{momdef4}). The convolution kernel is 
\be
D(z;m) = i\int\frac{d^2 p}{2\pi} \th(p_0) \delta(p^2 - m^2) \,
e^{-i p\cdot z}\;,
\label{Evol11}
\ee
which for real $z$ would coincide with the free scalar two-point 
function of mass $m$. In (\ref{Evol10}) the argument is given by
\ba 
&& z^+ = i(q^{-1/2} x^+ + q^{1/2} y^+)\;,\;\;\; 
z^- = -i(q^{1/2} x^- + q^{-1/2} y^-)\;,\;\;\;\mbox{or}\nonum
&& z^{\mu} = x^{\mu}(\tau) - y^{\mu}(-\tau)\bigg|_{\tau = 
\frac{i\pi}{2}(1 - \frac{2\pi}{\beta})}\;,
\label{Evol12}
\ea
in the normalization (\ref{I3}). Before specializing to the 
particular complex arguments (\ref{Evol12}) let us recall a few basic
facts on the analyticity properties of $D(z;m)$ in the complex 
domain in general. First, since the measure in (\ref{Evol11}) has support 
only in the (open) forward lightcone $V^+$, $D(x;m)$, $x \in \R^{1,1}$, 
is the boundary value of an analytic function holomorphic in the 
forward tube $\{z \in \C^2\,|\, - {\rm Im}z \in V^+\}$.
This function admits a further analytic continuation due to the fact that
(\ref{Evol11}) is invariant even under complex Lorentz transformations. 
This implies $D(z;m) = F(z^2)$, where 
$F(w)$ is holomorphic in the cut plane $\C\backslash \R^+$. Indeed, 
evaluating the integral (\ref{Evol11}) for $z$ in the forward tube 
yields
\be
D(z;m) = \frac{i}{2\pi}K_0(m \sqrt{- z^2})\;,
\label{Evol13}
\ee
where we use $z^2 = (z^0)^2 - (z^1)^2 = 2 z^+ z^-$ also for $z \in \C^2$ 
and $K_0$ is a modified Bessel function. The excluded region where 
(\ref{Evol13}) fails is where $z^2$ is real and non-negative.
For $\beta =2\pi$ this is the case whenever the separation of $x,y$ is 
timelike or null, so that one recovers the familiar `Euclidean' 
behavior of the free two-point function at spacelike distances
$(x - y)^2 <0$. For $\beta \neq 2\pi$ (or rather $\g \neq 0$)
$z^2$ is never real unless $x^+ y^- - y^+ x^- =0$, in which case 
$z^2$ is negative iff $x$ and $y$ are spacelike separated.
In other words for $\beta \neq 2\pi$ (\ref{Evol13}) holds iff 
\ba
A(x,y) \neq 0\;,\;\;\;\mbox{or}\;\;\; A(x,y) =0\;\;\mbox{with}\;\;
x,y\;\;\mbox{spacelike}\;,
\label{Evol14}
\ea 
where
\be
A(x,y) := \frac{1}{2}\det{ x^0\;\;y^0 \choose x^1\;\;y^1} = 
A(x(\lb),y(\lb))\;.
\label{Adef}
\ee 
Geometrically $A(x,y)$ is the (oriented) area enclosed by the 
three points $x,y,0$, if $0$ denotes the `origin' of $W$. It is also
closely related to the central extension of the 1+1 dim. Poincar\'{e}    
group $P_+^{\uparrow}$. Indeed if $g_1 = (a_1,\lb_1)$, $g_2 = (a_2,\lb_2)$ 
are two elements of $P_+^{\uparrow}$ parameterized by a translation
parameter $a$ and a boost parameter $\lb$ then $\omega(g_1,g_2) :=
A(a_1, a_2(\lb_1))$ is a 2-cocycle for  $P_+^{\uparrow}$.  
The cocycle already appeared in other contexts in low dimensional
quantum geometry \cite{Jackiw}.

It also reappears when we consider now the 
response of the two-point function (\ref{Evol10})
with respect to a variation of $\beta$. It is useful to generally denote 
the first ${\scriptstyle \beta}\dd{\beta}$ derivative of some quantity 
evaluated at $\beta =2\pi$ by a subscript $R$ (for ``response'').
In particular for the free scalar two-point function we set
\be
D_R(x,y;m) := \beta \dd{\beta} D(z;m)\bigg|_{\beta = 2\pi} =
\frac{i\pi}{2} \dd{\tau} 
D\Big(x(\tau) - y(-\tau);m\Big)\bigg|_{\tau = 0}\;.
\label{Evol15}
\ee  
This function has a number of interesting properties. First, as is clear 
from the second expression, it has again an interpretation within the context
of Minkowski space QFT. It describes the response of a two-point function  
upon Lorentz boosting the points $x,y$ by an infinitesimal oppositely 
equal amount. Taking into account (\ref{Evol13}) one obtains
\be
D_R(x,y;m) = -\frac{m}{4}\frac{A(x,y)}{\|x -y\|} 
K_1(m\|x -y\|)\;,\;\;\;x,y\;\;\mbox{spacelike}\;,
\label{Evol16}
\ee
employing $\|x\| := \sqrt{-x^2}$ for $x^2 <0$ and  
$$
\beta \dd{\beta}\, z^2\bigg|_{\beta = 2\pi} = 
-i\pi \dd{\tau} x(\tau)\cdot y(-\tau)\bigg|_{\tau =0} =
-i \pi A(x,y)\;.
$$
In particular, in contrast to $D(x-y;m)$ itself, $D_R(x,y;m)$ has a 
well-defined scaling limit
\be
\lim_{\lb \ra 0^+} D_R(\lb x,\lb y;m) = \frac{1}{4} 
\frac{A(x,y)}{\|x-y\|^2}\;,\;\;\;x,y\;\;\mbox{spacelike}\;.
\ee 
Returning to the interacting case the response of a two-point
function (\ref{Evol10}) is given by 
\ba
W_R(x,y) \is \int_0^{\infty} d\mu \,\rho_R(\mu)\, (i\partial_+)^{2l_+}
(i\partial_-)^{2l_-} D(x-y;\mu)
\nonum 
&\!+\!& \int_0^{\infty} d\mu \,\rho(\mu)\, (i\partial_+)^{2l_+}
(i\partial_-)^{2l_-} D_R(x,y;\mu)\;,  
\ea
where $\rho(\mu) = \sum_{n \geq 1}\rho^{(n)}(\mu)$. Note that 
for the response $\rho_R(\mu)$ of the spectral density only $n \geq 2$ 
intermediate particles contribute. The simplest example for $\rho(\mu)$ 
having a non-vanishing (quadratic) response occurs for the energy 
momentum tensor of a free theory. 

\newsubsection{Energy momentum tensor: Form factors and free spectral
densities}

\begin{figure}[htb]
\vspace{3mm}
\mbox{$\makebox[3.5cm]{ } m\rho(s m)$}
\vskip 2mm
\leavevmode
\hskip 30mm
\epsfxsize=100mm
\epsfysize=60mm
\epsfbox{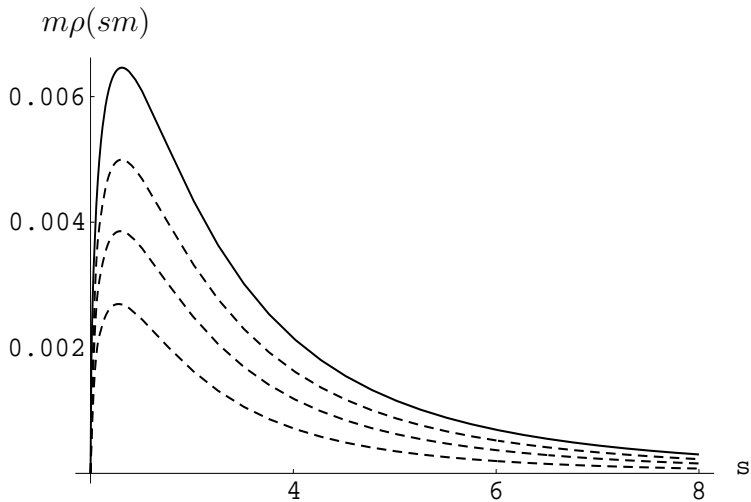}
\caption{EM spectral density for a free Majorana fermion. Undeformed
(solid) and deformed (dashed) for $\g = 0.5,\,0.7,\,0.9$, in order of 
decreasing maxima.}
\label{RhoF}
\end{figure}

Due to the conservation equation the form factors of the 
energy momentum (EM) tensor provide a link between kinematical
and dynamical aspects of a theory. Here let us denote by 
$F^{(n)}_{\mu\nu}(\th)$ a deformed counterpart of the EM 
form factors. As noted in section 2 such a counterpart
should always exist and it can be assumed to transform according to 
(\ref{deff6}). Possible ambiguities in its definition can 
be constrained by by means of the conservation equation. 
In the undeformed case the conservation equation 
$P^{(n)}(\th)^{\mu}\,F^{(n)}_{\mu\nu}(\th) =0$ of course reflects 
the Poincar\'{e} invariance of the underlying QFT. 
Here we cannot presuppose such a framework.
Nevertheless it turns out that the conservation equation
\be
P^{(n)}_+(\th) F^{(n)}_{- \pm}(\th) + 
P^{(n)}_-(\th) F^{(n)}_{+ \pm}(\th) =0\;,
\label{EM1}
\ee
can be consistently imposed in the following way. Beginning with 
$F^{(2)}_{+-}(\th_{21})$ regular at $\th_{21} = \pm i\pi$ and normalized 
according to $F_{+-}^{(2)}(i\beta/2)=m^2$ there will exist a unique 
sequence $F_{+-}^{(n)}(\th)$ of deformed form factors satisfying  
a suitable minimality condition (c.f. the Sinh-Gordon model below for 
an exemplification). The definition
\be
F^{(n)}_{\pm \pm}(\th) := 
-\left(\frac{P_+^{(n)}(\th)}{P_-^{(n)}(\th)}\right)^{\pm 1}\,
F^{(n)}_{+ -}(\th)\;,
\label{EM2}
\ee
then supplements the other two components. Note that this
definition does not exclude that other solutions for 
$F_{\pm \pm}^{(n)}(\th)$ exist for which (\ref{EM1}) doesn't hold.
Adopting (\ref{EM2}), however, equation (\ref{EM1}) holds and it 
follows that all components of $F_{\mu \nu}^{(n)}(\th)$ can be 
parameterized in terms of the deformed momentum eigenvalues and a 
boost invariant ``scalarized'' form factor $f^{(n)}(\th)$
\be
F^{(n)}_{\pm \pm}(\th) = -P^{(n)}_{\pm}(\th)^2\,
f^{(n)}(\th)\;,
\sspace 
F^{(n)}_{\pm \mp}(\th) = P_+^{(n)}(\th)P_-^{(n)}(\th)\,
f^{(n)}(\th)\;.
\label{EM3}
\ee
The simplest examples are that of a free boson and a free Majorana 
fermion of mass $m$, where only the two-particle EM form factor is 
nonvanishing. The deformed scalarized EM form factors are  
\ba
\mbox{Free Boson:} \sspace
f^{(2)}(u) \is \frac{2m^2}{M^{(2)}(u)^2}\;,
\nonum
\mbox{Free Fermion:} \sspace 
f^{(2)}(u) \is -i\frac{2m^2}{M^{(2)}(u)^2}\,\sh\frac{u}{2}\;,
\label{EM4}
\ea
where 
\be
[M^{(2)}(u)/m]^2 = \frac{2(1 +\ch u)(2 - \g^2 + 2 \ch u)^2}%
{4 + \g^4 + 4(2-\g^2) \ch u + 4(1-\g^2)\ch^2 u}\;. 
\ee
Using (\ref{diff6}) the deformed spectral densities (\ref{Evol10})
can be computed and are functions of $\g^2 = (2\cos\pi^2/\beta)^2$ 
only. One finds 
\ba
\mbox{Free Boson:}\;\;\;m \rho(\mu) \is \frac{2}{\pi}
\frac{1}{s^4\sqrt{s^2 -4}}\;,\sspace\forall \g^2 <1\;,\nonum
\mbox{Free Fermion:}\;\;\;m \rho(\mu) \is \frac{1}{2\pi}
\frac{\sqrt{s^2 -4}}{s^4}\left( 1 - \g^2 +O(\g^4)\right)\;,
\label{EM6}
\ea
for $s = \mu/m \geq 2$. Remarkably the EM spectral density of 
the free boson is not affected by the deformation while that of the 
free fermion is. This feature can be understood from the fermionic 
S-matrix $S=-1$, which, though still a phase, is typical for an 
interacting theory in 1+1 dimensions. (Recall that generically bootstrap 
S-matrices for interacting QFTs satisfy $S_{ab}^{cd}(0) = -\d_a^d\d_b^c$.)
Technically this forces the form factors to have a zero at $u =0$
and thus to be qualitatively different from those for the trivial S-matrix
$S=1$. Observe also that the $O(\g^2)$ correction to the 
fermionic spectral density is negative. This is not compensated  
by the subleading terms. Solving $M^{(2)}(u) = \mu/m$ for 
$\ch u$ one encounters a cubic equation, so that 
the explicit evaluation of the deformed spectral densities is conveniently 
done numerically. The result for various values of $\g$ is shown in 
Fig.~\ref{RhoF}. 

With hindsight the decrease of $\rho(\mu)$ is not counterintuitive, 
keeping in mind that $m\rho(\mu)$ is a dimensionless measure for the number 
of mass-weighted degrees of freedom coupling to the EM tensor at energy 
$\mu$. Since the mass eigenvalues were increasing for $\beta$ off $2\pi$ 
one expects this number to decrease, at least for the free case. 

The central charge is naturally defined to be the 
coefficient of the $1/(z^+)^4$ singularity in the EM two-point
function, with the normalization fixed by the undeformed case. 
This amounts to 
\be
c(\g) = 12 \pi \int_0^{\infty}\,d\mu \rho(\mu)\;,
\label{EM7}
\ee
where we anticipated that (here) it is an (even) function of $\g$ 
only. From (\ref{EM6}) one has $c(\g) \equiv 1$ for the bosonic
case while for the fermion $c(\g)$ is monotonously decreasing 
in $|\g|$
\be
\mbox{Free Fermion:}\;\;\;c(\g) = \frac{1}{2}(1 - \g^2 + O(\g^4))\;,
\sspace c(\pm 1) = 0.128(1)\;.
\label{EM8}
\ee
Incidentally the effect on the central charge of taking $\beta$ off 
(but close to) $2\pi$ is the same as adding a background charge to the 
Lagrangian. After bosonizing the fermion in terms of a bose field $\phi$, 
for example a curvature term $\sim \gamma R \phi$ could account for
this. The flat space EM tensor then received a correction $\sim \gamma
\Box \phi$, giving rise to a non-vanishing 1-particle form factor
$F^{(1)}_{+-} \sim \gamma$ in the undeformed theory. In this way 
the $O(\g^2)$ term in (\ref{EM8}) could be mimicked in an ordinary 
$\beta = 2\pi$ QFT. Note however that the trace of the energy momentum 
tensor still has a vanishing expectation value, i.e.~$F^{(0)}_{+-} = 0$
in terms of form factors.

In conformally invariant theories the quantity $\int d^2 x \bra 
\Theta(x)\ket_{\beta}$ has been computed by different techniques    
\cite{Dow,HLW,MorV} and found to be proportional to 
$c\,(1 - (\frac{2\pi}{\beta})^2)\frac{\beta}{2\pi}$, where $c$ is 
the central charge of the CFT and $\Theta$ is the trace of the EM 
tensor. The non-zero result there is due 
to the fact that in CFT a different definition of the EM tensor is
used: As explained in \cite{CFL} scale invariant theories have 
a spectral density supported at zero $\rho(\mu) \sim c\, \d(\mu)$.
Inserting this into the Euclidean version of the spectral representation
(\ref{Evol10}) of the $T_{++}$ two-point function yields a contact term
$\bra T_{++}(x) T_{++}(0)\ket \sim c \,\partial^2 \d^{(2)}(x)$. 
Such contact terms can be modified by adding local terms to the
effective action, i.e.~their form depends on the renormalization
scheme. In CFT one uses a scheme where the contact term is removed 
at the expense of (the new) $T_{++}$ no longer transforming as a true
tensor. Rather the transformation law involves the well-known Schwarzian 
connection. Using the fact that the Schwarzian of the mapping
$x^{\pm} \ra  (x^{\pm})^{\beta/2\pi}$ is 
$\frac{1}{2}(1 - (\frac{2\pi}{\beta})^2)$ one readily obtains the 
quoted result for $\int d^2 x \bra \Theta(x)\ket_{\beta}$ \cite{HLW}.
In genuinely massive theories however there is no reason 
to redefine the EM tensor in that way. One is then lead to 
impose $F_{+-}^{(0)} \equiv 0$ also for the $\beta \neq 2\pi$ 
systems and arrives at the results (\ref{EM4}), (\ref{EM6}) 
(\ref{EM8}). An important aspect of (\ref{EM8}) is that the
central charge changes at all. Expanding $\g^2$ in a way that 
preserves the $\beta \ra -\beta$ invariance gives $\g^2 = 
\frac{\pi^2}{4}(1 - (\frac{2\pi}{\beta})^2)^2 + \ldots $.
Thus the change in the central charge is subleading as compared to the 
CFT change in $\int d^2 x \bra \Theta(x)\ket_{\beta}$, -- which 
explains why it hasn't been seen in \cite{Dow,HLW,Mor,MorV}.

Also for generic interacting QFTs we expect the central charge 
to depend on $\beta$. This then indicates that the $\beta \neq 2\pi$ 
deformation does in general not commute with a conventional 
renormalization group transformation. The latter would otherwise just 
`squeeze' the initial $\beta =2\pi$ spectral density without changing 
the area enclosed by its graph, i.e.~the central charge \cite{CFL}.

\newsection{Form factors in the deformed Ising and 
Sinh-Gordon model}

As an illustration for the deformation procedure for interacting QFTs
we present here a few sample form factors in the deformed counterparts 
of the Ising model and the Sinh-Gordon model. Both models have been 
extensively studied from the viewpoint of form factors. They have a 
scalar diagonal S-matrix and three soliton super-selection sectors: 
A bosonic, a fermionic and a disorder sector, reflecting an underlying 
$Z_2$-symmetry. Some major references in the context of form factors are 
\cite{Ising1,Ising2a,Ising2b,Ising3,Ising4,Ising5} for the Ising model and 
\cite{SinhG1,SinhG2,SinhG3,SinhG4,SinhG5} for the Sinh-Gordon theory.

\newsubsection{Ising model} 

We restrict attention to the form factors of the 
spin field $\sigma$ and the disorder field $\mu$. Set%
\footnote{The prefactor $c^{(n)}=(2 i)^{[n/2]}$ is fixed up to a real 
overall constant by the residue equation $(i/2){\rm res}F^{(n)} = 
[\eta (-)^{n-2} -1]F^{(n-2)}$ and hermiticity, resulting in the 
conditions $c^{(n)} = 2 i\, c^{(n-2)}$, $[c^{(n)}]^* = 
(-)^{n(n-1)/2}\,c^{(n)}$, respectively.}
\be
f^{(n)}(\th) = (2 i)^{[n/2]}\,\prod_{k>l}\tanh\frac{\th_{kl}}{2}\;,
\label{Ising1}
\ee
where $[x]$ denotes the integer part of $x \in \R$. Then 
$f^{(n)}(\th)$ is the $n$-particle form factor of $\sigma/\mu$ for
$n$ odd/even, respectively \cite{Ising1,Ising2a,Ising2b}. On parity grounds 
the even/odd form factors of $\sigma/\mu$ vanish. In the deformed case 
the defining relations for the form factors of the Ising model are 
($\eta =1$ for $\sigma$ and $\eta =-1$ for $\mu$)
\ba
&& f^{(n)}(\th_n,\th_{n-1},\ldots \th_1) =-
f^{(n)}(\th_{n-1},\th_n,\th_{n-2},\ldots,\th_1)\;,\nonum
&& f^{(n)}(\th_n+i\beta,\th_{n-1},\ldots,\th_1) = 
f^{(n)}(\th)\;,\nonum
&& \mbox{res}_{\th_{n}=\th_{n-1} \pm i\pi}f^{(n)}(\th)=
\frac{i\beta}{\pi}\,f^{(n-2)}(\th_{n-2},\ldots,\th_1)\;.
\label{Ising3}
\ea
An appropriate Ansatz for the deformed spin and disorder form 
factors is
\ba
&& f^{(n)}(\th_n,\ldots, \th_1) = g^{(n)}(t_n,\ldots,t_1)\; 
(2 i)^{[n/2]}\prod_{k>l} 
\frac{t_k -t_l}{(t_k - q t_l)(t_k -q\inv t_l)}\;,
\label{Ising4}
\ea
where  $g^{(n)}(t)$ is a symmetric polynomial in 
$t_j = e^{2\pi\th_j/\beta}$. Inserting the Ansatz (\ref{Ising4}) into 
the deformed residue equations (\ref{Ising3}) yields 
the recursive relations 
\ba
&\nspace & g^{(n)}(q^{\pm 1}t_{n-1},t_{n-1},\ldots,t_1)\nonum
&\nspace & = t_{n-1}(1+q^{\pm 1}) 
\prod_{k=1}^{n-2} (t_{n-1} -q^{\pm 1}t_k)%
(q^{\pm 1}t_{n-1} -q^{\mp 1}t_k)\;
g^{(n-2)}(t_{n-2},\ldots,t_1)\;.
\label{Ising5}
\ea 
Starting with $g^{(0)}=1= g^{(1)}$ there exists a unique polynomial 
solution $(g^{(n)}(t))_{n\geq 0}$ with 
$g^{(n)}(t) \in P^{(n)}(n(n-1)/2, n-1)$, which reduces to 
$\prod_{k>l}(t_k +t_l)$ in the limit $\beta \ra 2\pi$. 
In fact these solutions happen to coincide with the denominators
of the conserved charges $P^{(n)}(t)$ described in section 3, i.e.
\be
g^{(n)}(t) = \den P^{(n)}(t)\;,\;\;\forall n \geq 0\;.
\label{Ising6}
\ee
In particular for $n\leq 4$ the explicit expressions are already listed 
in (\ref{charge7}). An explanation for the coincidence (\ref{Ising6}) 
is given in the Appendix. The same polynomials once more reappear in 
the form factors of the deformed Sinh-Gordon model. 

\newsubsection{Sinh-Gordon model} 

The undeformed form factors for the Sinh-Gordon
model are likewise well-known \cite{SinhG1,SinhG2}. An appropriate 
Ansatz for the deformed form factors turns out to be  
\be
F^{(n)}(\th) = c^{(n)}h^{(n)}(\th)\;g^{(n)}(t)
\prod_{k > l} \frac{\psi(\th_{kl})}{(t_k - q t_l)(t_k -q\inv t_l)}\;,
\label{SHG1}
\ee
where the $c^{(n)}$ are constants, $t_j = e^{2\pi\th_j/\beta}$, as 
before and $g^{(n)}(t)$ are the ``Ising model'' polynomials 
(\ref{Ising6}) solving (\ref{Ising5}). The functions to be determined 
are $h^{(n)}(\th)$, which are completely symmetric 
and $i\beta$-periodic in all variables. Finally $\psi(u)$ is the deformed 
minimal form factor, solving $\psi(u) = S(u) \psi(-u)$
and $\psi(u+i\beta) = \psi(-u)$. The solution analytic in the strip 
$0 \leq {\rm Im}\,u < \beta/2$ is given by
\be
\psi(u) = - i \cN\,\sh\frac{\pi u}{\beta} 
\exp\left\{ -2 \int_0^{\infty} \frac{ds}{s} 
\frac{\ch \frac{(B-1)\pi}{\beta}s}{\ch\frac{\pi}{\beta}s\,\sh s}\;
\sin^2 \frac{s}{2\pi}\Big(i\pi - \frac{2\pi}{\beta}u\Big) \right\}\;.
\label{SHG2}
\ee 
As indicated, it has a simple zero at $u =0$ and no others in the strip
of analyticity. The normalization constant $\cN$ is real and is chosen 
such that $\psi(u) \ra 1$ for $u \ra \pm \infty$. Further $0<B<2$ is the 
effective coupling constant, transforming as $B \ra 2 -B$ under the 
weak-strong coupling duality. In these conventions the Sinh-Gordon 
S-matrix reads 
$S(\th) = (\sh\th -i\sin\frac{\pi}{2} B)/(\sh\th +i\sin\frac{\pi}{2} B)$
and is invariant under the duality transformation.%
\footnote{We assume here that $\pi B/\beta$ is irrational, though 
interesting resonance phenomena might occur by fine-tuning $\beta$
and the coupling.}
Entering with the Ansatz  (\ref{SHG1}) into the deformed form factor 
equations only the residue equation remains and becomes
\ba
&\nspace & h^{(n)}(\th_{n-1} \pm i\pi,\th_{n-1},\ldots ,\th_1) = 
\pm 2 i\,\frac{t_{n-1}(q^{\pm 1}-1)}{\psi(i\pi)}\,\frac{c^{(n-2)}}{c^{(n)}}
\times \nonum
&\nspace & \times \prod_{k=1}^{n-2} 
2 i q^{\pm 1/2} \,t_{n-1}t_k \,\Upsilon(\pm \th_{n-1,k})
\;h^{(n-2)}(\th_{n-2},\ldots,\th_1)\;.
\label{SHG3}
\ea 
The function 
\be
\Upsilon(u) := -2 i
\frac{\sh\frac{\pi}{\beta}u \,\sh\frac{\pi}{\beta}(u +i\pi)}%
{\psi(u)\psi(u + i\pi)}
\label{SHG4}
\ee
can easily be seen to have the following properties. It is a smooth 
function on $\R$ approaching $-\frac{iq}{2}e^{2\pi |u|/\beta}$ at real 
infinity. It is $i\beta$-periodic and obeys the functional equations   
$\Upsilon(-u) = \Upsilon(u -i\pi)$, $\Upsilon(u)^* = -\Upsilon(-u^*)$. 
For $\beta = 2\pi$ it simplifies to 
$\Upsilon(u) = \sh u +i\sin\frac{\pi}{2} B$. For generic $\beta$ an explicit 
evaluation is more cumbersome but can still be achieved
\be
\Upsilon(u) = i\cos\frac{\pi^2}{\beta}(1-B) -i\,\ch\bigg(\frac{2\pi}{\beta}u 
+\frac{i\pi^2}{\beta}\bigg)\;.
\label{SHG5}
\ee
From here one anticipates that also in the deformed Sinh-Gordon model
the computation of form factors can be reduced to a polynomial problem.
Indeed, introducing the definitions 
\ba
&& D^{(n-2)}(x;t_{n-2},\ldots ,t_1) =
-x\prod_{k =1}^{n-2}(x - \omega t_k)(x - \omega^{-1} t_k) 
\nonum
&& = -x\left[ \sum_{l=0}^{n-2} x^{2(n-2 -l)}\sigma_l^2
+ \sum^{n-2}_{l > k\geq 0} x^{2n - 4 -k -l}(-)^{l+k}[\omega^{l-k} + 
\omega^{-(l-k)}]\;\sigma_l \sigma_k\right]\;,
\label{SHG6}
\ea
(where $\sigma_l = \sigma_l^{(n-2)}(t),\,l=0,\ldots, n-2$) and
\be
\omega := e^{-\frac{i\pi^2}{\beta}(1-B)}\;,
\sspace c^{(n)}:=c^{(n_0)} \left(
\frac{4 \sin\frac{\pi^2}{\beta}}{\psi(i\pi)}\right)^{\frac{n-n_0}{2}}
\;\;,\;n \geq n_0\;,
\label{SHG7}
\ee
the relation (\ref{SHG3}) translates into 
\be
h^{(n)}(q^{\pm 1} t,t,t_{n-2},\ldots ,t_1) =
D^{(n-2)}(q^{\pm 1/2} t;t_{n-2},\ldots ,t_1) \;
h^{(n-2)}(t_{n-2},\ldots ,t_1)\;.
\label{SHG8}
\ee
Here $n_0$ refers to the starting member of a sequence, the square
root of $q$ is $q^{1/2} = e^{i\pi^2/\beta}$, and we write $h^{(n)}(t)$ for 
$h^{(n)}(\th)$. In the form (\ref{SHG8}) the solutions for low $n$ are readily 
found. However, provided one insists on having the proper (polynomial) 
$\beta \ra 2\pi$ limit the relevant solutions turn out to be ratios 
of symmetric polynomials. 

For definiteness we restrict attention to the form factors of the 
elementary field $\phi$ and the EM tensor. Their $n$-particle
form factors are denoted by $F^{(n)}(\th)$ and $F^{(n)}_{\mu\nu}(\th)$,
respectively. As in the undeformed case we stipulate that the even 
particle form factors of $\phi$ vanish and that the odd particle 
form factors of the EM tensor vanish. The normalization conditions are
\be
F^{(1)}(\th) =1\;,\sspace F^{(2)}_{+-}(\th) = 
\frac{m^2}{\psi(\frac{i\beta}{2})}\psi(\th_{21})\;.
\label{SHG9}
\ee
Referring to the Ansatz (\ref{SHG1}) it is convenient to set 
$F^{(n)}(\th) := F^{(n)}_{+-}(\th)$ for $n$ even. 
For the constants $c^{(n_0)}$ in (\ref{SHG7}) this means $c^{(1)} =1$
and $c^{(2)} = m^2/\psi(i\beta/2)$. With these conventions the 
functions $h^{(n)}(t)$ are found to be of the form 
\ba
&\nspace & \bspace \bspace\;\;\;\;
h^{(n)} = \frac{\num h^{(n)}}{\den h^{(n)}}\;,
\nonum
&\nspace & \num h^{(n)} \in P^{(n)}(n(n-1), 2 n -2)\;,   
\;\;\;\;\den h^{(n)} \in P^{(n)}\bigg(\frac{n(n-1)}{2}, n-1\bigg)\;.
\label{SHG10}
\ea
Using the shorthand $A = 2 \cos\frac{\pi^2}{\beta}(1-B)$ the 
explicit expressions for $n \leq 4$ are $h^{(1)}=1$ and
\ba
&\nspace & \num h^{(2)} = \s_1^2 - \g^2 \s_2\;, 
\sspace  \den h^{(2)} = \s_1\;,
\nonum
&\nspace & \num h^{(3)} = A \s_1\s_2\s_3 - A \g(\s_2^3 + \s_1^3 \s_3) +
A [\g A^2 + A(-\g + \g^2(5 + \d)) \nonum
&& \sspace -3 - \d + 2\g(5 + 3\d)]\,\s_3^2\;,
\nonum
&\nspace & \den h^{(3)} = 
\s_1\s_2 + ((-1+A)\g + 7 + 6\d + \d^2)\,\s_3\;,
\nonum
&\nspace & \num h^{(4)} = A(\s_1 \s_2 \s_3)^2 - \g(1+ A \g)
(\s_2^3 \s_3^2 + \s_1^3\s_3^3 + \s_1^2\s_2^3\s_4)
\nonum 
&&\sspace + (A(5 + 5\d + \d^2)+ 2 \g^3)%
(\s_1 \s_2 \s_3^3 + \s_1^3 \s_2 \s_3 \s_4)
\nonum 
&&\sspace -\g(2 + A \g)(1 + \d)^2(\s_3^4 + \s_1^4 \s_4^2) 
+ \g^3(2 + A \g)\,\s_2^4 \s_4 -\g R_1 \s_1 \s_2^2 \s_3 \s_4 
\nonum
&& \sspace - \g(A^2 + A \g\d - 2 + \d)\,\s_1^2 \s_3^2 \s_4 
+ \g R_2\, (\s_2 \s_3^2 \s_4 + \s_1^2 \s_2 \s_4^2) 
\nonum
&& \sspace - \g \d R_3\, \s_1 \s_3 \s_4^2
- \g^3 \d (A^2 +2A \g(1 + \d) + 2 + 5\d)\,(\s_2\s_4)^2 
\nonum
&&\sspace 
+ \g^3 \d^3 (A^2 + A \g^3 + 2 + 3\d)\,\s_4^3\;,
\nonum
&\nspace & \den h^{(4)} = \s_1\s_2\s_3 -(1 + \d)^2 (\s_1^2\s_4 + \s_3^2)
+\g^4\d\,\s_2\s_4\;.
\label{SHG11}
\ea
The shorthands are 
\ba
R_1 \is  A \g(6 + 6\d + \d^2) + (12 + 14\d + 3\d^2)\;,
\nonum
R_2 \is A^2(1 + \d) + A\g(2 + 8\d + 6\d^2 + \d^3)
+(2 + 15 \d + 14\d^2  + 3 \d^3)\;
\nonum
R_3 \is  2A^2 + A \g^3(2 + 2 \d + \d^2) + 
(4+ 14\d + 10\d^2 + 3\d^3)\;.
\label{SHG12}
\ea 
For $\beta \ra 2\pi$ these expressions reduce to the polynomials
\be
h^{(2)} \ra \s_1\;,\;\;\;\;h^{(3)} \ra 2 \sin\frac{\pi}{2}B\,\s_3\;,
\;\;\;\; h^{(4)} \ra  2 \sin\frac{\pi}{2}B\,\s_1\s_2\s_3\;,
\ee
so that (\ref{SHG1}) gives back the undeformed form factors. 
The expressions (\ref{SHG11}) are the minimal deformed counterparts 
in the sense that they are ratios of symmetric polynomials with
the smallest possible degrees. Each solution could be modified
by adding a solution of (\ref{SHG8}) vanishing in the limit 
$\beta \ra 2\pi$. The expressions (\ref{SHG11}) are also minimal 
in the sense that such additions have been omitted. In the EM case
one might be tempted to use an Ansatz of the form $h^{(n)}(t) = 
P^{(n)}(t) \bar{h}^{(n)}(t)$, with some remainder $\bar{h}^{(n)}(t)$ again
in the form of a ratio of symmetric polynomials. However the resulting
solutions would have much higher degrees as in (\ref{SHG10}) and 
thus would be non-minimal in the above sense.  
Finally notice that the coefficients in (\ref{SHG12}) are real which 
ensures hermiticity $h^{(n)}(t)^* = h^{(n)}(t^*)$. 

\newsection{Conclusions}

We have implemented the replica notion of taking the Unruh
temperature $\beta$ off its physical value $2\pi$ for a large class
of interacting QFTs. The technique developed allows one to compute
the response of a QFT with a factorized scattering operator under a
variation of $\beta$. It automatically produces finite cutoff-independent 
answers for these response functions and in principle can be 
applied to any local quantum field theoretical quantity one might be 
interested in. We will comment on bulk quantities below. 

Among the physically notable results is the increase (\ref{momdef5}) 
of the mass eigenvalues on asymptotic states. This means it
costs more energy to boost two particles relative to each other
than in the undeformed case. The cost function has the form of a 
plateau cut by steep valleys along the diagonals of the rapidity phase 
space, i.e. $\R^n$ for $n$ particles. Configurations where two or 
more particles asymptotically move parallel will therefore give the dominant 
contributions to phase space integrals. Nevertheless as long as 
$\g = 2\cos \pi^2/\beta$ is less than unity the height of the plateau 
is finite and the entire rapidity phase space remains accessible. 
This ceases to hold as one crosses the $\g=1$ 
barrier. For example for $n=2$ the positivity condition (\ref{momdef3}) 
can be seen to put an upper bound on the relative rapidity of the two 
particles. Generally only part of the original phase space is accessible 
for $\g \geq 1$ and the regions excluded are those with extremely high 
relative boost parameters.
This is very much in the spirit of 't Hooft's picture of scattering
states subject to quantum gravitational `transmutation' \cite{tHooft}. 
The idea is that each individual particle can be Lorentz boosted 
arbitrarily. However relative boosts of two or more particles  
corresponding to trans-Planckian energies should be `transmuted' 
into cis-Planckian ones in a way dictated by the formalism.  
The formalism should take into account the fluctuating horizon. 
Here we mimicked such fluctuations by varying the quantity conjugate 
to its area. Of course in `t Hooft's  picture also the directions
transverse to the Rindler horizon (absent here) play a decisive role.
It should be interesting to see whether, -- after extending (\ref{ffcycl}), 
(\ref{I5}) to higher dimensions -- similar patters emerge from the present
framework. 
 
We also found that the central charge of the systems will in general
depend on $\beta$, indicating that the $\beta \neq 2\pi$ deformation
does not commute with ordinary renormalization group transformations. 
One will also be interested in other bulk quantities like the free energy 
and the relative entropy. A natural framework to compute them in the present 
context is the thermodynamic Bethe Ansatz. Although we kept the 
bootstrap S-matrix fixed the relevant integral equation may be 
modified nevertheless for $\beta \neq 2\pi$. Since the integral equation
can be derived from the form factor approach \cite{Balog} one 
can in principle work out the modified integral equation and
compute bulk quantities for the $\beta \neq 2\pi$ systems. First
the free energy and then through its $\scriptstyle{\beta}\dd{\beta}$
response the entanglement entropy \cite{GeoSa,GeoSb}; see 
\cite{KSS} for a perturbative treatment in the O(N) model.   
It is also tempting to ask whether the $\beta \neq 2\pi$ systems 
introduced here can arise as the continuum limit of some (novel) 
statistical mechanics systems. 

Finally it should be worthwhile to examine the geometrical aspects in 
more detail. Here we concentrated mainly on momentum space issues. 
The position space geometry of the deformed systems, in particular the 
extent to which deformed versions of translation invariance 
and micro-causality exist, remains to be explored.

\newpage
\setcounter{section}{0}

\newappendix{Solution of recursive equations}

Here we collect some details on the solution of recursive relations 
of the form
\be
G^{(n)}(q^{\pm 1}t,t,t_{n-2}, \ldots,t_1)
= D^{(n-2)}(q^{\pm 1/2}t;t_{n-2}, \ldots,t_1) \, 
G^{(n-2)}(t_{n-2},\ldots,t_1)\;.
\label{A1}
\ee 
The $G^{(k)}(t)$ are symmetric functions in $t_j = e^{2\pi\th_j/\beta}$,
$j =1,\ldots, k$ and $D^{(k)}(x;t)$ is a polynomial in $x$ whose
coefficients are symmetric polynomials in $t_1,\ldots,t_k$.
Recursive relations of this type appeared at three 
different instances in the bulk of the paper: (i) In the definition of 
the conserved charges, equation (\ref{charge3}) with  $D^{(k)}(x;t)=1$.
(ii) In the Ising model, equation (\ref{Ising5}) with $D^{(k)}(x;t)$ 
given explicitly below. (iii) In the Sinh-Gordon model, equations
(\ref{SHG6}), (\ref{SHG8}). The solutions searched for are ratios
of symmetric polynomials in $t_1,\ldots,t_k$ with a prescribed 
$\beta \ra 2\pi$ limit. Provided also the degrees of the numerator
and denominator polynomials are taken to be the smallest possible the 
solutions turn out to be uniquely specified by these requirements 
up to trivial ambiguities.   

In preparation let $P^{(n)}(N,l)$ denote the space of homogeneous 
symmetric polynomials in $t_1,\ldots,t_n$ of total degree $N$ and
partial degree $l$ (where the partial degree is the maximal degree 
in an individual variable). Let $\nu  =(\nu_1, \ldots,\nu_l)$,
$\nu_1 \geq \ldots \geq \nu_l \geq 0$ be a partition of $N$ into 
$l$ parts less or equal $n$, i.e. $\sum_i \nu_i = N$, $0\leq \nu_i \leq n$,
$1 \leq i\leq l$. Running through all these partitions, the assignment
\be
(\nu_1, \ldots,\nu_l) \rra \sigma_{\nu_1}^{(n)}\ldots    
\sigma_{\nu_l}^{(n)}
\label{pol2}
\ee
provides a basis of  $P^{(n)}(N,l)$, where 
\be
\sigma_k^{(n)} =\sum_{i_1<\ldots <i_k} t_{i_1}\ldots t_{i_k}\;,
\sspace k=1,\ldots, n\;,
\label{pol1}
\ee
are the elementary symmetric polynomials. The reduction operation 
$t_n \ra  q^{\pm 1} t_{n-1}$ entering (\ref{deff1}) takes the 
form
\be
\sigma_k^{(n)} \rra  \sigma_k^{(n-2)} +(1 + q^{\pm 1}) t_{n-1}
\sigma_{k-1}^{(n-2)} + q^{\pm 1} t_{n-1}^2 \sigma_{k-2}^{(n-2)}\;,
\label{pol3}
\ee
with $\sigma_k^{(n-2)} = 0$ for $k <0$ or $k > n-2$. The simultaneous
sign flip of all the rapidities $t_j \ra t_j^{-1}$ becomes
$\sigma_k^{(n)} \ra  \sigma_{n-k}^{(n)}/\sigma_n^{(n)} =: 
{\overline \sigma}^{(n)}_k$. For later use let us also note that 
the reduction operation (\ref{pol3}) has a kernel which can be 
described as follows. Set
\be
E^{(n)}(t) = \prod_{j<k}(t_j-q t_k)(t_j-q^{-1} t_k) =
(\sigma^{(n)}_1\sigma^{(n)}_2 \ldots \sigma^{(n)}_{n-1})^2 + \ldots \;.
\label{pol5}
\ee
The dots indicate subleading terms with $q$-dependent coefficients. 
Clearly $E^{(n)}(t) \in P^{(n)}(n(n-1), 2n-2)$ and lies in the kernel of the 
reduction operation $t_n \ra q^{\pm 1} t_{n-1}$. Further, it is the 
element of the kernel with the smallest total degree, it is the only element 
with this total degree, and all other polynomial elements of the kernel 
are obtained by multiplying $E^{(n)}(t)$ with a symmetric polynomial. 

With these preparations let us consider (\ref{A1}) with 
\ba
&& D^{(n-2)}(x;t_{n-2},\ldots ,t_1) = 
\g x\prod_{k =1}^{n-2}(x - q^{3/2}t_k)(x - q^{-3/2} t_k) 
\nonum
&& = \g x\left[ \sum_{l=0}^{n-2} x^{2(n-2 -l)}\sigma_l^2
+ \sum^{n-2}_{l > k\geq 0} x^{2n - 4 -k -l}(-)^{k+l} [q^{3(l-k)/2} 
+ q^{-3(l-k)/2}]\;\sigma_l \sigma_k\right]\;,
\label{A2}
\ea
where $\sigma_l = \sigma_l^{(n-2)}(t),\,l=0,\ldots, n-2$. 
This is relevant for two situations. First the Ising model, where 
(\ref{Ising5}) is of the form (\ref{A1}) with the above $D$'s.
Second it turns out that the numerators and denominators of 
the $s =1$ power sums $P^{(n)}(t)$ separately satisfy (\ref{A1}) with
the $D$'s given by (\ref{A2}). More specifically one finds that 
(\ref{A1}), (\ref{A2}) admits a unique sequence of solutions 
$G^{(n)}_s(t)$, $s=0,1$ obeying
\be
G^{(n)}_s(t) \in P^{(n)}(n(n-1)/2 + s, n-1 +s) \;,\;\;\;
G^{(n)}_s(t)\bigg|_{\beta = 2\pi} = 
(t_n^s + \ldots + t_1^s)\prod_{k>l}(t_k + t_l)\;.
\label{A3}
\ee
Moreover by construction their ratio solves (\ref{charge3}) 
and in facts meets all the requirements in the definition of 
the $s=1$ deformed power sums. Whence 
\be
G^{(n)}_1(t) = \num P^{(n)}(t)\;,\;\;\;
G^{(n)}_0(t) = \den P^{(n)}(t)\;.
\label{A4}
\ee
In particular table (\ref{charge7}) also provides the $n\leq 4$ 
members of the $s=0,1$ solutions to (\ref{A1}), (\ref{A2}), (\ref{A3}). 
For $s\geq 3$ this construction no longer works (e.g. it fails 
for $s=3$ and $n=4$). However one may consider (\ref{A1}) with 
\be
D^{(n-2)}(x;t_{n-2},\ldots ,t_1) = 
\left[\g x\prod_{k =1}^{n-2}(x - q^{3/2}t_k)
(x - q^{-3/2} t_k)\right]^p\;,
\label{A5}
\ee
i.e. with the right hand side of (\ref{A2}) raised to some power $p$.
Of course a trivial way to produce solutions of this recursive 
relation is to raise some $p =1$ solution to its $p$-th power.
However there are also solutions which are not of this form.
In fact the power sum eigenvalues discussed in section 3 are precisely 
nontrivial solutions of (\ref{A1}), (\ref{A5}) in this sense. 
If we momentarily denote by ${}^pG_s^{(n)}(t)$ a 
solution of (\ref{A1}) with $D^{(k)}$ given by (\ref{A5}) and 
$\beta \ra 2\pi$ limit $(t_1^s +\ldots + t_n^s)\prod_{l>k}(t_k + t_l)^p$, 
then 
\be
\num P_s^{(n)} = {}^{2s}G_s^{(n)}(t)\;,\;\;\;
\den P_s^{(n)} = [G_0^{(n)}(t)]^{2s}\;,\;\;\mbox{for}\;\;s>0\;,
\label{A6}
\ee
while for $s <0$ the roles of the numerator and denominator are 
interchanged. Clearly the degrees of the numerator and denominator 
polynomials will usually be fairly large and one may often find solutions  
with smaller degrees, which otherwise meet the same requirements.
In contrast to the undeformed case moreover not any such quantity 
can be obtained as a product or ratio of power sums. In other words
for $\beta \neq 2\pi$ the power sums do not provide a basis for the 
ring of conserved charge eigenvalues described in section 3.1.   
An explicit counterexample in given in equation (\ref{pol10}), 
(\ref{pol11}) below.

Generally speaking the point is that a solution of recursive 
equations of the form (\ref{A1}) is not uniquely specified by its 
$\beta \ra 2\pi$ limit. In order to uniquely specify a solution additional
requirements have to be imposed. A trivial ambiguity arises from
the spin zero conserved charges $Q_0^{(n)}(t)$. This is because 
any $n$-particle solution of the deformed form factor equations can always be 
multiplied with $(1+ \g^2 Q_0^{(n)})$ without affecting
the properties under the reduction operation $\th_n =\th_{n-1} \pm i\pi$, 
its spin, or the $\beta \ra 2\pi$ limit. Solutions from which one cannot 
split off such a factor might be called ``primary''. But also the primary 
solutions are not uniquely determined by their $\beta \ra 2\pi$ limit. 
In the bulk of the paper we considered solutions of the deformed 
form factor equation which (possibly after splitting off a universal 
transcendental piece) were ratios of symmetric polynomials. For such 
solutions it is natural to choose the 
solutions where the numerator and denominator have the smallest 
possible degrees. In all the cases considered we found that this
additional requirement fixed the solution up to trivial ambiguities.
Depending on the context however also other requirements may be natural.
An example is the definition (\ref{momdef2}), (\ref{momdef4}) of 
the deformed momentum and mass eigenvalues. There we insisted 
on having $P_+^{(n)}(t) \sim \sum_j k_j^{(n)}(t)$ and 
$P_-^{(n)}(t) \sim \sum_j [k_j^{(n)}(t)]^{-1}$ with the {\em same}
functions $k_j^{(n)}(t)$ in both cases. This lead to the 
non-primary expressions (\ref{momdef2}). Lorentz invariance then 
enforces to take (\ref{momdef4}) as the definition of the 
deformed mass eigenvalues. These are obviously again non-primary,
but even after splitting off $Q^{(n)}(t)$ the remainder 
$P^{(n)}(t)P^{(n)}(t\inv)$ is not the solution with the smallest
possible degrees of the numerator and denominator polynomials.

We conclude this appendix by giving the explicit expressions for 
the minimal solution. It also provides an example for a 
spin zero conserved charge that cannot be expressed as a product 
or ratio of power sums. It is the minimal primary spin zero
conserved charge having $\sigma_1^{(n)}{\overline \sigma}_1^{(n)}$ as 
its $\beta \ra 2\pi$ limit and will be denoted by $Q_0^{(n)}(t)$ below. 
We already encountered two other spin zero conserved 
charges having the same $\beta \ra 2\pi$ limit, namely 
$M^{(n)}(t)$ as defined in (\ref{momdef4}) and $P^{(n)}(t)P^{(n)}(t\inv)$.
However $M^{(n)}(t)$ is not primary and $P^{(n)}(t)P^{(n)}(t\inv)$ not
minimal in the above sense. The latter can be anticipated by noting 
that the element $E^{(n)}(t)$ in the kernel of the reduction operation
(\ref{A4}) is of total degree $n(n-1)$, less than that of the product 
$P^{(n)}(t)  P^{(n)}(t\inv)$. The structure of the minimal 
spin zero conserved charge with $\sigma_1^{(n)}{\overline \sigma}_1^{(n)}$ as 
its $\beta \ra 2\pi$ limit can be described as follows:
\ba
&\nspace & \bspace \bspace\;\;\;\;
Q_0^{(n)} = \frac{\num Q_0^{(n)}}{\den Q_0^{(n)}}\;,
\nonum
&\nspace & \num Q_0^{(n)} \in P^{(n)}(n(n-1), 2 n -2)\;,   
\;\;\;\;\den Q_0^{(n)} \in P^{(n)}(n(n-1), 2n-3)\;.
\label{pol10}
\ea
Since the degrees of the numerator coincide with that of $E^{(n)}(\lb)$
the solution is unique only up to addition of a multiple of it
vanishing in the $\beta \ra 2\pi$ limit. This trivial ambiguity can be 
fixed by requiring that the numerator contains $(\s_1\s_2\ldots\s_{n-1})^2$
with unit coefficient, c.f. (\ref{pol5}). With these specifications 
the solution is unique and for $n \leq 4$ the explicit expressions are 
given by 
\ba
&& Q_0^{(2)} = \frac{E^{(2)}}{\s_2}\;,\sspace 
E^{(2)} = \s_1^2 - \g^2 \s_2\;, 
\nonum 
&& Q_0^{(3)} = 1 + \frac{E^{(3)}}{\s_3(\s_1 \s_2 - \s_3)}\;,\nonum
&& E^{(3)} = (\s_1\s_2)^2 - \g^2(\s_2^3 + \s_1^3\s_3) + 
(1 + \d)(4 +\d)\s_1\s_2\s_3 - (1 + \d)^3 \s_3^2\;,
\nonum
&& \num Q_0^{(4)} = (\s_1\s_2\s_3)^2 - \s_1^3\s_2\s_3\s_4 - 
\d \g^2(1+ \d)^2\,\s_1^4 \s_4^2 
- \s_1\s_2\s_3^3 - \g^4\, \s_2^4\s_4 \nonum 
&& \;\;\;+ 2\g^2(1 +\d)\, \s_1\s_2^2 \s_3 \s_4 -
\d \g^2(1+ \d)^2\,\s_3^4 - \d^2 \g^4(1 + 2\d)\, \s_2^2 \s_4^2 \nonum
&& \;\;\; + \d\g^2(1 + 6\d + 5\d^2 + \d^3)\, 
( \s_2\s_3^2 \s_4 + \s_1^2 \s_2\s_4^2) \nonum
&& \;\;\; - 2\d^2 \g^2(1 + \d)(2 + 2\d + \d^2)\,
\s_1 \s_3 \s_4^2 + 2\d^4 \g^4 (1 + \d)\, \s_4^3\nonum 
&& \den Q_0^{(4)}/\s_4 = \g^2\,\s_2^4 - (3 + 2\d) \,\s_1\s_2^2\s_3 - 
\g^4\d\,\s_1^2\s_3^2 \nonum
&& \;\;\; + (1 + \d)(1 + 4\d + 2\d^2)\,
(\s_2\s_3^2 +\s_1^2\s_2\s_4) - \d \g^2 (4 +6\d + \d^2) \, 
\s_2^2\s_4 \nonum
&& \;\;\; + \g^2\d(2 + 2 \d + 3\d^2 + \d^3)\,\s_1\s_3 \s_4 
- \g^2\d^3 (2 + 2\d + \d^2)\,\s_4^2\;.
\label{pol11}
\ea




\newpage
\begin{thebibliography}{10}
\bibitem{Ising4} O. Babelon and D. Bernard, \PL{B288} (1992) 113.
\bibitem{Balog} J. Balog, \NP{B419} (1994) 480.
\bibitem{Ising1} B. Berg, M. Karowski and P. Weisz, \PR{D19} (1979) 2477.
\bibitem{BW} J. Bisognano and E. Wichmann, 
J. Math. Phys. {\bf 16} (1975) 985; 
J. Math. Phys. {\bf 17} (1976) 303; 
\bibitem{GeoSa} L. Bombelli, R. Kaul, J. Lee and S. Sorkin,
\PR{D34} (1986) 373.
\bibitem{BTZ} M. Bonadis, C. Teitelboim and J. Zanelli, \PRL{72} 
(1994) 957.
\bibitem{CW} C. Callan and F. Wilczek, \PL{B333} (1994) 55.
\bibitem{CFL} A. Capelli, D. Friedan and J. Latorre,
\NP{B352} (1991) 616.
\bibitem{Ising5} J. Cardy and G. Mussardo, \NP{B410} (1993) 451.
\bibitem{Dow} J. Dowker, \PR{D36} (1987) 3095.
\bibitem{SinhG1} A. Fring, G. Mussardo and P. Simonetti, \NP{B393} 
(1993) 413. 
\bibitem{FR} S. Fulling and S. Ruijsenaars, Phys. Rep. {\bf 152} (1987)
135.
\bibitem{HLW} C. Holzhey, F. Larsen and F. Wilczek, \NP{B424}
(1994) 443.
\bibitem{Jackiw} R. Jackiw: Higher symmetries in lower dimensional
models, Proceedings Salamanca, 1992.
\bibitem{KSS} D. Kabat, S. Shenker and M. Strassler,
\PR{D52} (1995) 7027.
\bibitem{KS} D. Kabat and M. Strassler, \PL{B329} (1994) 46.
\bibitem{SinhG2} A. Koubek and G. Mussardo,  \PL{\bf B311} 
(1993) 193. 
\bibitem{SinhG5} V. Korepin and N.A. Slavnov,
The determinant representation for quantum correlation functions 
of the Sinh-Gordon model, hep-th/9801046.
\bibitem{SinhG3} M. Lashkevich, Sectors of mutually local fields in 
integrable models of QFT, hep-th/9406118.
\bibitem{Ising2b} E. Marino, B. Schroer and J. Swieca, \NP{B200} (1982) 473.
\bibitem{Mor} E. Moreira, \NP{B451} (1995) 365.
\bibitem{MorV} V. Moretti and L. Vanco, \PL{B375} (1996) 54.\\
V. Moretti, Class. Quant. Grav. {\bf 14} (1997) L123.
\bibitem{MNmod} M. Niedermaier, \NP{B519} (1998) 517. 
\bibitem{MNcycl} M. Niedermaier, \CMP{196} (1998) 411.
\bibitem{SinhG4} M. Pillin, The form factors in the Sinh-Gordon 
model, hep-th/9712033.
\bibitem{Ising2a} B. Schroer and T. Truong, \NP{B144} (1978) 80.
\bibitem{Smir} F.A. Smirnov, {\em Form Factors in Completely Integrable 
Models of QFT}, World Scientific, 1992.
\bibitem{GeoSb} M. Srednicki, \PRL{71} (1993) 666. 
\bibitem{tHooftS} G. 't Hooft, Int. J. Mod. Phys. {\bf A11} (1996) 4623.
\bibitem{tHooft} G. 't Hooft: Trans-Planckian particles and the 
quantization of time, gr-qc/9805079.
\bibitem{Unruh} W. Unruh, Phys. Rev. {\bf D14} (1976) 870.
\bibitem{Ising3} V. Yurov and Al.B. Zamolodchikov, Int. J. Mod. 
Phys. {\bf A6} (1991) 3419.
\end{thebibliography}
\end{document}